\documentstyle[11pt]{article}\textheight 230mm\textwidth 150mm
            \newcommand{\be}{\begin{eqnarray}}
            \newcommand{\ee}{\end{eqnarray}}
            \newcommand{\eel}[1]{\label{#1}\end{eqnarray}}
\newcommand{\e}[1]{\label{e:#1}\end{eqnarray}}
     \newcommand{\eg}{{\em e.g.\ }}
            \newcommand{\ie}{{\em i.e.\ }}
            \newcommand{\ga}{{\gamma}}
 
            \newcommand{\la}{{\lambda}}
            
            \newcommand{\del}{{\delta}}
           \newcommand{\ra}{{\rightarrow}}

            \newcommand{\pet}{{\cal P}}
 \newcommand{\tpet}{\tilde{\cal P}}
 \newcommand{\tca}{\tilde{{\cal C}}}
 \newcommand{\tpe}{\tilde{p}}

\newcommand{\ca}{{\cal C}}
\newcommand{\baca}{\bar{\cal C}}
            
            \newcommand{\beq}{\begin{quote}}
            \newcommand{\eq}{\end{quote}}
            
            \newcommand{\al}{\alpha}
            \newcommand{\ben}{\begin{enumerate}}
            \newcommand{\een}{\end{enumerate}}
            \newcommand{\bit}{\begin{itemize}}
            \newcommand{\ei}{\end{itemize}}
    	\newcommand{\nn}{\nonumber}
            \newcommand{\r}[1]{(\ref{e:#1})}
            \newcommand{\edfl}[1]{\label{#1}\end{df}}

\newcommand{\vb}{{\cal h}}
\newcommand{\hb}{{\cal i}}

\newcommand{\th}{{\theta}}
\newcommand{\bapet}{\bar{\pet}}
\newcommand{\dagg}{^{\dag}}
\newcommand{\sign}{\mbox{sign}}
\newcommand{\bett}{{\bf 1}}
\newcommand{\halv}{\frac{1}{2}}
	
\begin{document}
\begin{titlepage}
\noindent
G\"{o}teborg ITP 95-13\\
June 1995\\
hep-th/9507023\\

\vspace*{5 mm}
\vspace*{35mm}
\begin{center}{\LARGE\bf Gauge fixing and coBRST}\end{center}
\vspace*{3 mm} \begin{center} \vspace*{3 mm}

\begin{center}G\'eza F\"ul\"op\footnote{E-mail: geza@fy.chalmers.se} and Robert
Marnelius\footnote{E-mail: tferm@fy.chalmers.se} \\ \vspace*{7 mm}
{\sl Institute of Theoretical Physics\\ Chalmers University of Technology\\
G\"{o}teborg University\\
S-412 96  G\"{o}teborg, Sweden}\end{center}
\vspace*{25 mm}
\begin{abstract}
It has previously been shown that a BRST quantization on
an inner product space leads
to physical states of the form $$|ph\hb=e^{[Q,
\psi]} |\phi\hb$$ where  $|\phi\hb$   is either
a trivially BRST invariant state
which only depends on the matter  variables,
$|\phi\hb_1$, or a solution of a Dirac
quantization, $|\phi\hb_2$.
$\psi$ is a corresponding fermionic gauge
fixing operator, $\psi_1$ or $\psi_2$.
We show here for  abelian and nonabelian
models that one may also choose a linear combination
of $\psi_1$ and $\psi_2$ for
both choices of $|\phi\hb$ except for a discrete set of
relations between the coefficients. A general
form of the coBRST charge operator is also
determined and shown to be equal to such a $\psi$ for an allowed linear
combination of $\psi_1$ and $\psi_2$. This means that the
coBRST charge is always a
good  gauge fixing fermion. \end{abstract}\end{center}\end{titlepage}

\setcounter{page}{1}
\setcounter{equation}{0}
\section{Introduction.}
In a BRST quantization one starts from a BRST invariant theory
defined on a nondegenerate inner product
space, $V$, and projects out the BRST invariant states. Of particular
interest are the BRST singlets, $|s\hb$, which represent the true
physical states ($|s\hb\in \mbox{Ker} Q/\mbox{Im} Q$). They may be
chosen to be orthogonal to all unphysical
states in $V$. In ref.\cite{Solv} it was shown that within the operator
formulation of
general BRST invariant theories with finite number of degrees of freedom
the BRST
singlet states on inner product spaces may be represented in the form
\be &&|s\hb=e^{[Q, \psi]}|\phi\hb \e{1}
where $Q$ is the hermitian, nilpotent BRST charge, $\psi$ a hermitian fermionic
gauge fixing operator and $|\phi\hb$ a BRST invariant state vector which does
not belong to an inner product space. More precisely it was shown that
there exist
two sets of hermitian operators each consisting of BRST doublets in involution,
\ie we
have \be
&&D_{(l)r}=\{C_{(l)i}, B_{(l)i}\},\;\;\;l=1,2
\e{2}
where (discarding factors of $i$)
\be
&&B_{(l)i}=[Q, C_{(l)i}]
\e{200}
which satisfy
\be
&&[D_{(l)r}, D_{(l)s}]={K_{(l)rs}}^tD_{(l)t}=
D_{(l)t}{K^{\dag\;\;t}_{(l)sr}},\;\;\;l=1,2
\e{3}
(We use graded commutators \cite{Solv}.) In addition, these sets also satisfy
the condition
\be
&&[D_{(1)r}, D_{(2)s}]\; \mbox{ is an invertible matrix operator.}
\e{4}
The doublets \r{2} determine two equivalent forms of formula \r{1}, namely
\be
&|s\hb_l=e^{[Q, \psi_l]}|\phi\hb_l,\;\;\;l=1,2
\e{5}
where the states $|\phi\hb_l$ satisfy the conditions
\be
&&D_{(l)r}|\phi\hb_l=0,\;\;\;l=1,2
\e{6}
which are consistent due to \r{3}, and which
always imply that $|\phi\hb_l$ are BRST invariant.
For the states $|s\hb_l$ we have then
\be
&&D'_{(l)r}|s\hb_l=0,\;\;\;l=1,2
\e{06}
where
\be
&&D'_{(l)}\equiv e^{[Q, \psi_l]}D_{(l)}e^{-[Q, \psi_l]}
\e{07}
If
\be
&&[D'_{(l)r}, (D'_{(l)s})\dagg]\; \mbox{ is an invertible matrix operator}
\e{08}
then $D'_{(l)r}$ and $(D'_{(l)s})^{\dag}$ constitute BRST quartets
and  $|s\hb_l$ are singlet states due to the quartet mechanism \cite{KO,Nish}.
Condition \r{08} determines the allowed class of
 hermitian gauge fixing fermions $\psi_l$ in \r{5}.
In \cite{Solv} it was shown that
an allowed choice is
 \be
&&\psi_1=C^{(b)}_{(2)a}C^{(f)a}_{(2)},\;\;\;
\psi_2=C^{(b)}_{(1)a}C^{(f)a}_{(1)}
\e{7}
where $C^{(b)a}$ and $C^{(f)a}$ are the bosons and  fermions respectively
of the $C$-operators in
the BRST doublets \r{2} which are assumed to commute in \r{7}.  (Notice that
the condition  \r{4} requires the
$C$-operators to consist of equally many bosons and fermions.)

When these results  were applied to general, both irreducible and reducible
gauge theories of arbitrary rank within the BFV formulation in
\cite{Solv} it was shown that
 there always exists a rather
simple representation of the two sets of BRST doublets $D_{(l)}$ which makes
\r{6} simple to solve.
 For instance, for an arbitrary irreducible gauge theory we have the
BFV-BRST charge operator
\be
&&Q=\ca^a\th_a+\bapet^a\pi_a+\ldots
\e{8}
where $\ca^a$ are ghost operators,  $\bapet^a$ conjugate momenta to the
antighosts $\baca^a$, and $\pi_a$
 the conjugate momenta to the Lagrange multipliers $ v^a$. $\th_a$ are
the gauge generators
which are in involution. The dots in \r{8} represent terms involving the
ghosts $\ca^a$ and their
conjugate momenta $\pet_a$. They are determined by the condition $Q^2=0$
and the precise form of the
commutator $[\th_a, \th_b]$ (see \eg refs \cite{BF1,BF2}). (For abelian
gauge theories,
$[\th_a, \th_b]=0$, the first two terms in \r{8} are sufficient.) In this
case the BRST
doublets, $D_{(l)}$, are naturally given by (apart from factors of $i$)
\cite{Solv}
\be
&&D_{(1)}=\{\chi^a, [Q, \chi^a]; \baca_a, \pi_a\}\nn\\
&&D_{(2)}=\{ v^a, \bapet^a; \pet_a, [Q, \pet_a]=\th_a+\ldots\}
\e{9}
where $\chi^a$ are gauge fixing conditions to $\th_a$ which are required
to be in involution.
Conditions \r{6} imply then that $|\phi\hb_1$ is a ghost fixed state which does
not
depend on the Lagrange multipliers and the gauge generators, while $|\phi\hb_2$
satisfies a Dirac quantization
apart from also being a ghost fixed state which does not depend on the Lagrange
multipliers.  The corresponding gauge fixing fermions  may be chosen to be
\be
&&\psi_1=\pet_a v^a,\;\;\;\psi_2=\baca_a\chi^a
\e{10}
One may notice that the two  singlets
\be
&&|s\hb_l=e^{[Q, \psi_l]}|\phi\hb_l,\;\;\;l=1,2
\e{11}
are BRST invariant states even if $|\phi\hb_l$ does not satisfy the following
gauge fixing conditions  \be
&&\chi^a|\phi\hb_1=0,\;\;\;  v^a |\phi\hb_2=0
\e{12}
In
ref.\cite{Simple,Proper,Gauge}
the expressions \r{11} without the conditions \r{12} were obtained
by means of a bigrading in the
case when the gauge group is  a general Lie group.

In this paper we present two generalizations of the results of refs
\cite{Solv,Simple,Proper,Gauge}. First we consider the possibility to
generalize the form of the gauge fixing fermions \r{10}.
Remember that in the conventional treatment
of the BFV-theory the gauge fixing fermions, which there enter into the
Hamiltonians, are usually chosen to be a linear combination of $\psi_1$
and $\psi_2$ in \r{10} (see
\eg ref. \cite{BF1}). Since these gauge fixing fermions are equal to ours
apart from
a multiplicative time parameter according to ref.\cite{Path} we should
also be able to use
 gauge fixing fermions which are  linear combinations of $\psi_1$ and $\psi_2$
in \r{11}. Indeed, in sections  3 and 6 we prove  for
 abelian respectively nonabelian models that \r{5}   are still singlet states
as well as inner product
states when $\psi_l$ is an arbitrary linear combination of those in \r{7}  and
when  $|\phi\hb_l$ satisfy \r{6} {\em except for a discrete set of relations
between  the coefficients of $\psi_1$ and $\psi_2$}.

The second subject of this paper concerns  the role of coBRST invariance in
the above construction. In ref \cite{Solv} it was suggested that the coBRST
charge
operator, $^*Q$, should provide for a more invariant formulation. Indeed, the
conditions \be
&&Q|s\hb=\,^*Q|s\hb=0
\e{13}
do project out singlet states from an original nondegenerate inner product
space V
\cite{Nish,Spi,RRy,KvH}.   The coBRST charge is defined by
\be
&&^*Q\equiv \eta Q\eta
\e{14}
where $\eta$ is a hermitian metric operator which maps the original state
space V onto a Hilbert space. It satisfies
 \be
&&\vb u|\eta|u\hb\geq 0,\;\;\forall |u\hb\in V,\;\;\;\eta^2=\bett\nn\\
&&(\vb u|\eta|u\hb=0\;\Leftrightarrow\; |u\hb=0)
\e{15}
The coBRST charge $^*Q$ is simply the hermitian conjugate of $Q$ in this
Hilbert
space. Notice that $^*Q$ is also  nilpotent. In terms of the coBRST charge
we have
the Hodge decomposition, which means that any state $|u\hb\in V$ may uniquely
be
written as \be
&&|u\hb=|s\hb+Q|u_1\hb+\,^*Q|u_2\hb,
\e{16}
where  the singlet states $|s\hb$ are  determined by  \r{13} or equivalently
\be
&&\triangle|s\hb=0,\;\;\;\triangle\equiv[Q,\,^*Q]
\e{17}
(see \eg \cite{KvH}).
One may also show that every state in the non-physical space can be written as
a
linear combination of eigenstates of the $\triangle$ operator. The eigenvalues
corresponding to these eigenstates are positive real numbers. ($\triangle$ is
hermitian in the Hilbert space: $^*\triangle=\eta\triangle\eta=\triangle$.)

In section 2 we summarize known results for abelian models and in section 3
we derive
the conditions  for singlet states of the form \r{5} with generalized gauge
fixing
fermions.
 In section 4  we connect the form \r{5} with a general Fock space
construction,
and in section 5 we construct $\eta$ and  the coBRST charge $^*Q$ for the
simple  abelian model of sections 2 and 3. The BFV
form of $^*Q$ turns out to have the form of  an allowed
gauge fixing fermion $\psi$.
The relation between the gauge fixing fermion in \r{5} and the coBRST charge
operator that annihilates this state is also given.
In section 6 these results are generalized
to a class of nonabelian models. In section 7 we then summarize our results
and give
some concluding remarks. In three appendices we give  proofs of  formulas
and some  unitary transformations used in the text.

\setcounter{equation}{0}
\section{A simple abelian model}
In the following  we shall make extensive use of a very simple abelian model
whose hermitian BRST charge operator is given by
\be
&&Q=\ca^ap_a+\bapet^a\pi_a
\e{21}
where $p_a$ and $\pi_a$ are hermitian conjugate momenta to the
hermitian coordinates $x^a$ and
$ v^a$ respectively, and $\ca^a$ and $\bapet^a$ are hermitian
fermionic operators conjugate to
$\pet_a$ and $\baca_a$ respectively. The index $a=1,\ldots,n < \infty$
is assumed to be
raised and lowered by a real symmetric metric $g_{ab}$. The fundamental nonzero
commutators are \be &&[x^a, p_a]_-=i\del^a_b,\;\;\;[ v^a,
\pi_b]_-=i\del^a_b,\;\;\;[\ca^a, \pet_b]_+=\del^a_b,\;\;\;[\baca^a,
\bapet_b]_+=\del^a_b
\e{22}
One may think of \r{21} as the BRST charge operator of an abelian
bosonic gauge theory where $p_a$
are the gauge generators, $ v^a$ the Lagrange multipliers, and $\ca^a$ and
$\baca_a$ the ghosts and
antighosts respectively. Alternatively one may view it as the BRST
charge of a fermionic gauge
theory with bosonic ghosts $p_a$ and antighosts $ v^a$, or a mixture of these
two interpretations.

Applying the rules of ref.\cite{Solv} as described in the introduction
we find here the
two dual sets $D_{(l)}$ in \r{2}  to be given by (cf \r{9})
\be
&&D_{(1)r}=\{x^a, \ca^a, \baca_a, \pi_a\},\;\;\;D_{(2)r}=\{ v^a, \bapet^a,
\pet_a, p_a\}
\e{23}
which obviously satisfy \r{3} and \r{4}. Thus, we obtain the singlet states
\be
&&|s\hb_l=e^{[Q, \psi_l]}|\phi\hb_l,\;\;\;l=1,2
\e{24}
where we may \eg choose the gauge fermions
\be
&&\psi_1=\al\pet_a v^a,\;\;\;\psi_2=\beta\baca_a x^a
\e{25}
for arbitrary finite nonzero real constants $\al$ and $\beta$,
and where $|\phi\hb_1$
and $|\phi\hb_2$ satisfy
\be
&&x^a|\phi\hb_1=\ca^a|\phi\hb_1=\baca_a|\phi\hb_1=\pi_a|\phi\hb_1=0\nn\\
&& v^a|\phi\hb_2=\bapet^a|\phi\hb_2=\pet_a|\phi\hb_2=p_a|\phi\hb_2=0
\e{26}
A more general allowed choice than \r{25} is
\be
&&\psi_1=\pet_aT^{ab} v^b,\;\;\;\psi_2=\baca_aS^{ab}x^b
\e{251}
where $T^{ab}$ and $S^{ab}$ are real, invertible matrices.

For \r{25} the
conditions
\r{26} imply
\be
&&D'_{(1)}|s\hb_1=0,\;\;\;D'_{(2)}|s\hb_2=0
\e{27}
where
\be
&&D'_{(1)}=\{x^a+i\al v^a, \ca^a+i\al\bapet^a, \baca_a-i\al\pet_a, \pi_a-i\al
p_a\}\nn\\
&&D'_{(2)}=\{ v^a+i\beta x^a, \bapet^a+i\beta \ca^a, \pet_a-i\beta\baca_a,
p_a-i\beta\pi_a\}
\e{29}
If $\al$ and $\beta$ are different from zero and finite in
\r{25}, the condition \r{08} is satisfied, \ie we have that
$[D'_{(l)r}, (D'_{(l)s})^{\dag}]$ is an invertible matrix operator
for both $l=1$ and $l=2$ and $|s\hb_l$ are singlet states. They are then also
inner
product states:   In \cite{Proper} it was explicitly shown that $_l\vb
s|s\hb_l=\,_l\vb \phi|e^{2[Q,\psi_l]}|\phi\hb_l$ are finite for the above
model if
$\al$ and $\beta$ are finite and non-zero. In fact, for the more general choice
\r{251} we have \be
&&_1\vb s|s\hb_1=\frac{\det T}{|\det T|}A,\;\;\;_2\vb s|s\hb_2=\frac{\det S}
{|\det S|}B
\e{301}
where $A$ and $B$ are finite expressions. (Thus, if $\al$ or $\beta$ is zero or
infinite in \r{25} we have the badly defined expressions $\vb s|s\hb=0/0$ or
$\vb
s|s\hb=\infty/\infty$!)

\setcounter{equation}{0}
\section{Generalized gauge fixing for the abelian model}
Here we  investigate under what conditions states of the form
\be
&&|s\hb_l=e^{[Q, \psi]}|\phi\hb_l,\;\;\;l=1,2
\e{31}
are singlet states for the abelian model where the gauge fixing fermion $\psi$
is a linear combination of those in \r{25}, \ie \be
&&\psi=\al\pet_a v^a+\beta\baca_a x^a
\e{32} (The states $|\phi\hb_1$
and $|\phi\hb_2$ are still required to satisfy  the conditions in \r{26}.)
 Thus, \r{31}
is a generalization of \r{24}. The conditions \r{26} imply now
\be
&&D'_{(l)r}|s\hb_l=0,\;\;\;l=1,2
\e{33}
where
$D'_{(l)}$  is given by \be
&&D'_{(l)}=e^{[Q, \psi]}D_{(l)}e^{-[Q, \psi]}
\e{34}
For $\al\beta>0$ we find $D'_{(1)r}$ to contain
\be
&&{x'}^a\equiv e^{[Q, \psi]}x^ae^{-[Q,
\psi]}=x^a\cos{\sqrt{\al\beta}}+i\al v^a\frac{\sin{\sqrt{\al\beta}}}
{\sqrt{\al\beta}}\nn\\
&&{\ca'}^a\equiv e^{[Q, \psi]}\ca^ae^{-[Q,
\psi]}=\ca^a\cos{\sqrt{\al\beta}}+i\al\bapet^a
\frac{\sin{\sqrt{\al\beta}}}{\sqrt{\al\beta}}\nn\\
&&{\baca'}_a\equiv e^{[Q, \psi]}\baca_ae^{-[Q,
\psi]}=\baca_a\cos{\sqrt{\al\beta}}-i\al\pet_a
\frac{\sin{\sqrt{\al\beta}}}{\sqrt{\al\beta}}\nn\\
&&{\pi'}_a\equiv e^{[Q, \psi]}\pi_ae^{-[Q, \psi]}=
\pi_a\cos{\sqrt{\al\beta}}-i\al
p_a\frac{\sin{\sqrt{\al\beta}}}{\sqrt{\al\beta}}
\e{35}
and $D'_{(2)r}$
\be
&&v'^a\equiv e^{[Q, \psi]} v^ae^{-[Q, \psi]}= v^a\cos{\sqrt{\al\beta}}
+i\beta x^a
\frac{\sin{\sqrt{\al\beta}}}{\sqrt{\al\beta}}\nn\\ &&{\bapet}^{'a}\equiv e^{[Q,
\psi]}\bapet^ae^{[-Q, \psi]}=\bapet^a\cos{\sqrt{\al\beta}}+i\beta
\ca^a\frac{\sin{\sqrt{\al\beta}}}{\sqrt{\al\beta}}\nn\\ &&{\pet'}_a\equiv
e^{[Q,
\psi]}\pet_ae^{-[Q,
\psi]}=\pet_a\cos{\sqrt{\al\beta}}-i\beta\baca_a
\frac{\sin{\sqrt{\al\beta}}}{\sqrt{\al\beta}}\nn\\
&&{p'}_a\equiv e^{[Q, \psi]}p_ae^{-[Q, \psi]}=p_a\cos{\sqrt{\al\beta}}
-i\beta\pi_a
\frac{\sin{\sqrt{\al\beta}}}{\sqrt{\al\beta}}
\e{36}
For $\al\beta<0$ we get the same
expressions with the replacements \be
&&\cos{\sqrt{\al\beta}}\,\ra\,\cosh{\sqrt{-\al\beta}},\;\;\;
\frac{\sin{\sqrt{\al\beta}}}{\sqrt{\al\beta}}\,\ra\,
\frac{\sinh{\sqrt{-\al\beta}}}{\sqrt{-\al\beta}}
\e{37}
In order to satisfy \r{08} we must have \be &&[x^{'a}, (\pi'_b)^{\dag}]_-
\mbox{ and } [\ca^{'a}, (\baca'_b)^{\dag}]_- \mbox{  are invertible}
 \e{38}
for $D'_{(1)r}$, and
\be
&&[v^{'a}, (p'_b)^{\dag}]_+ \mbox{ and } [\bapet^{'a},
(\pet'_b)^{\dag}]_+ \mbox{  are
invertible}
 \e{39}
for $D'_{(2)r}$. For \r{35} and \r{36} we find explicitly
\be
&&[x^{'a}, (\pi'_b)^{\dag}]_-=-\al
\frac{\sin{2\sqrt{\al\beta}}}{\sqrt{\al\beta}}\del^a_b,\;\;\; [\ca^{'a},
(\baca'_b)^{\dag}]_-=i\al
\frac{\sin{2\sqrt{\al\beta}}}{\sqrt{\al\beta}}\del^a_b\nn\\
&&[v^{'a},
(p'_b)^{\dag}]_+=-\beta
\frac{\sin{2\sqrt{\al\beta}}}{\sqrt{\al\beta}}\del^a_b,\;\;\;
[\bapet^{'a},
(\pet'_b)^{\dag}]_+=i\beta
\frac{\sin{2\sqrt{\al\beta}}}{\sqrt{\al\beta}}\del^a_b  \e{310}
Thus, for $\al\beta>0$ eq.\r{38} is satisfied provided $\al\neq 0$ and
$\sqrt{\al\beta}\neq n\frac{\pi}{2}$ where $n$ is a positive
integer, and eq.\r{39} is
satisfied provided $\beta\neq 0$ and $\sqrt{\al\beta}\neq n\frac{\pi}{2}$. For
$\al\beta<0$ we have to make the replacement \r{37} on the right-hand sides of
\r{310}. This implies that \r{38} and \r{39}   are
then satisfied for $\al\neq 0$ and
$\beta\neq 0$ respectively. This is true even in the limit $\al \beta\,\ra\,0$
in which case \r{35} and \r{36} reduce to \r{29}.

	We conclude that $|s\hb=e^{[Q, \psi]}|\phi\hb$ are singlet states
for the gauge fixing  \r{32} provided
$\al\neq 0$ ($\beta\neq 0$) when $|\phi\hb$ is chosen to satisfy the conditions
of $|\phi\hb_1$ ($|\phi\hb_2$) in \r{26}. In addition, we must have
$\sqrt{\al\beta}\neq
n\frac{\pi}{2}$ for any positive integer $n$ when $\al\beta>0$. In the path
integral formulation the conditions on $|\phi\hb$ correspond
to a choice of boundary
conditions \cite{Path}. Thus, when $\al\neq 0$, $\beta\neq 0$
and $\sqrt{\al\beta}\neq
n\frac{\pi}{2}$ one may choose any of the two sets of conditions
in  \r{26} as boundary
conditions.

It remains to investigate under which conditions the states \r{31} have finite
norms.
Let us write
\be
&&|s\hb_l=e^{\al K_1+\beta K_2}|\phi\hb_l,\;\;\;l=1,2
\e{321}
 where
we have introduced the hermitian operators
\be
&&K_1\equiv[Q, \pet_a v^a]= v^a p_a+i\pet_a\bapet^a\nn\\
&&K_2\equiv[Q, \baca_a x^a]=x^a \pi_a+i\baca_a \ca^a
\e{312}
If we in addition introduce the hermitian operator $K_3$ defined by
\be
&&K_3\equiv i\halv[K_1,
K_2]_-=\halv\left( v^a\pi_a-p_ax^a-i\pet_a\ca^a-i\bapet^a\baca_a\right)
=\nn\\
&&=\halv\left(\pi_a v^a-x^ap_a+i\ca^a\pet_a+i\baca_a\bapet^a\right)
\e{313}
we find that the algebra of the $K_i$ operators are  closed and given by
\be
&&[K_1, K_2]=-2iK_3,\;\;\;[K_1, K_3]=iK_1,\;\;\;[K_2, K_3]=-iK_2
\e{314}
This is an SL(2,R) algebra. (By means of the identification
$\phi_1=1/2(K_2-K_1),\;\;\phi_2=1/2(K_1+K_2),\;\;\phi_3=K_3$
we obtain the standard SL(2,R) algebra
$[\phi_i, \phi_j]=i\varepsilon_{ij}^{\;\;\;k}\phi_k
$
with the metric Diag($\eta_{ij})=(-1,+1,+1)$.)
By means of the properties
\be
&&K_2|\phi\hb_1=K_1  |\phi\hb_2=K_3    |\phi\hb_1=K_3|\phi\hb_2=0
\e{315}
it is then straight-forward to derive the following relations (a proof  is
given in appendix A) \be
|s\hb_1=e^{\al K_1+\beta K_2}|\phi\hb_1=e^{\al' K_1}|\phi\hb_1,\;\;\;
|s\hb_2=e^{\al K_1+\beta K_2}|\phi\hb_2=e^{\beta' K_2}|\phi\hb_2
\e{316}
where
\be
&&\al'=\al\,\frac{\tan{\sqrt{\al\beta}}}{\sqrt{\al\beta}},
\;\;\;\beta'=\beta\,\frac{\tan{\sqrt{\al\beta}}}{\sqrt{\al\beta}}
\e{317}
for $\al\beta>0$ and
\be
&&\al'=\al\,\frac{\tanh{\sqrt{-\al\beta}}}{\sqrt{-\al\beta}},
\;\;\;\beta'=\beta\,\frac{\tanh{\sqrt{-\al\beta}}}{\sqrt{-\al\beta}}
\e{318}
for $\al\beta<0$. Indeed \r{35} and \r{36} are equivalent to \r{29} with $\al$
and $\beta$ replaced by $\al'$
and $\beta'$. From \r{316} it follows that provided $\al'$
and $\beta'$ are non-zero
and finite  $|s\hb_1$ and $|s\hb_2$ are well defined inner product
states. The conditions  for this are  identical to the conditions
from \r{310} for $|s\hb_1$ and $|s\hb_2$ to be singlet states. Thus, as soon as
$|s\hb$ is a singlet state it is also an inner product state, and vice versa.

\setcounter{equation}{0}
\section{Fock space representation of the singlet states}
In order to acquire a deeper understanding of the results of section 3 we
construct here
 a  general Fock like representation of the singlet
states for the simple abelian model presented in
section 2. For this purpose we introduce the complex covariant bosonic
operators
\be
&&\phi_a\equiv ap_a+b\pi_a,\;\;\;\xi^a\equiv c v^a+dx^a
\e{401}
where $a,b,c$  and $d$ are complex constants. We require then
\be
&&[\xi^a, \phi_b]_-=0, \;\;\; [\xi^a, \phi_b^{\dag}]_-=\del^a_b
\e{402}
from which we  find
\be
&&c=\frac{ia}{ab^*-a^*b},\;\;\;d=\frac{ib}{a^*b-ab^*}
\e{404}
Similarly we  introduce the complex fermionic operators
\be
&&\rho^a\equiv e\ca^a+f\bapet^a,\;\;\;k_a\equiv g\pet_a+h\baca_a
\e{405}
The conditions
\be
&&[\rho^a, k_b]_+=0,\;\;\;[\rho^a, k_b^{\dag}]_+=\del^a_b
\e{406}
require here
\be
&&g=\frac{f}{fe^*-f^*e},\;\;\;h=\frac{e}{f^*e-fe^*}
\e{408}
We demand now that the BRST charge \r{21} must be possible to write as
\be
&&Q=\rho^{a\dag}\phi_a + \phi_a^{\dag}\rho^a
\e{409}
The reason is that a $Q$ of this form requires  the BRST invariant states,
which
contain  the singlet states, to satisfy the simple conditions
\be
&&\phi_a|ph\hb=\rho^a|ph\hb=0
\e{412}
or
\be
&&\phi^{\dag}_a|ph\hb=\rho^{a\dag}|ph\hb=0
\e{413}
For \r{401} and \r{405} the form \r{409} leads to the additional
condition \be
&&e=\frac{b}{ba^*-b^*a},\;\;\;f=\frac{a}{b^*a-ba^*}
\e{410}
which when inserted into \r{408} implies
\be
&&g=a,\;\;\;h=b
\e{411}
Our complex operators may  then be expressed in terms of only two arbitrary
constants $a$ and $b$ which are nonzero and subjected to the condition
$b^*a-ba^*\neq0$.

The corresponding singlet states to the physical states in \r{412} and
\r{413} satisfy
\be
&&\phi_a|s\hb=\rho^a|s\hb=k_a|s\hb=\xi^a|s\hb=0
\e{414}
or
\be
&&\phi^{\dag}_a|s\hb=\rho^{a\dag}|s\hb=k^{\dag}_a|s\hb=\xi^{a\dag}|s\hb=0
\e{415}
where the operators
$\phi_a,\;\rho^a,\;k_a,\;\xi^a,\;\phi^{\dag}_a,\;\rho^{a\dag},
\;k^{\dag}_a,\;\xi^{a\dag}$
constitute two sets of
 BRST quartets: ($\phi_a,\;\xi^{a\dag},\;
\rho^{a\dag},\;k_a$) and ($\phi^{\dag}_a,\;\xi,\;\rho^a,\;k^{\dag}_a$).
If there  are no other variables in the theory $|s\hb$
is just a vacuum state and all
the variables of the theory are unphysical.

 The question  is now whether or not the "vacuum" state $|s\hb$ defined by
\r{414}
or \r{415} is normalizable. In order to investigate this we make a
transition to a wave
function representation $\psi_s(\ca,\baca,x,v)=\vb \ca, \baca, x,v|s\hb$ where
$\ca^a,\; \baca_a,\; x^a$ and $ v^a$ are the eigenvalues of the corresponding
operators. The conditions \r{414} imply then \be &&0=\vb \ca, \baca,
x,v|\phi_a|s\hb=-i\left(a\frac{\partial}{\partial
x^a}+b\frac{\partial}{\partial
 v^a}  \right)\psi_s(\ca,\baca,x,v)\nn\\ &&0=\vb \ca, \baca,
x,v|\xi^a|s\hb=\left( c v^a+dx^a   \right)\psi_s(\ca,\baca,x,v) \e{416}
\be
&&0=\vb
\ca, \baca, x,v|\rho^a|s\hb=\left(e \ca^a+f\frac{\partial}{\partial \baca_a}
\right)\psi_s(\ca,\baca,x,v)\nn\\ &&0=\vb
\ca, \baca, x,v|k_a|s\hb=\left(a\frac{\partial}{\partial \ca^a}  +b\baca_a
\right)\psi_s(\ca,\baca,x,v) \e{417}
Obviously these conditions allow for solutions of the form
$\psi_s(\ca,\baca,x,v)=\psi_s(\ca,\baca)\psi_s(x,v)$
where \r{416} determines $\psi_s(x,v)$ and \r{417}  $\psi_s(\ca,\baca)$.
The solution of \r{416} is
\be
&&\psi_s(x,v)\propto\del^n(v+\frac{d}{c}x)
\e{418}
Now the argument of the delta function must be real. If \eg $ v^a$ and
$x^a$ have real
eigenvalues then
\be
&&\frac{e}{f}=\frac{d}{c}=-\frac{b}{a}
\e{419}
must be real. However, in this case we find
\be
&&\vb s|s\hb\propto\int d^nx d^n v \left(\del^n(v+\frac{d}{c}x)\right)^2
=\infty
\e{420}
On the other hand if one of the eigenvalues are imaginary we get
a finite result: Let
\eg $x^a$ have imaginary eigenvalues $iu^a$. The corresponding
eigenstates to $x^a$
satisfy then the relations \cite{Pauli,Gen}
\be
&&x^a|iu\hb=iu^a|iu\hb,\;\;\;\vb -iu|=(|iu\hb)^{\dag}\nn\\
&&\vb iu'|iu\hb=\del^m(u'-u),\;\;\;\int d^nu|-iu\hb\vb -iu|=\int d^nu|iu\hb\vb
iu|=\bett
\e{421}
which implies (in this case the ratio \r{419} must be imaginary
in order for the argument of the delta function
\r{418} to be real) \be
&&\vb s|s\hb\propto\int d^nu d^n v
\psi^*_s(-iu,v)\psi_s(iu,v)=\nn\\
&&
=\int d^nu
d^n v \del^n(v-i\frac{d}{c}u)\del^n(v+i\frac{d}{c}u) =
\left|\frac{c}{2d}\right|<\infty
\e{422} Similarly it follows that the bosonic part of $\vb s|s\hb$ is infinite
when
 both $x^a$
and $ v^a$ have imaginary eigenvalues, and that it is finite
also when $x^a$ is real
and $ v^a$ imaginary. For the fermionic part we get on the other hand zero for
\r{419} real, and finite when it is imaginary. The same results are obtained
if we
use \r{415} in \r{416} and \r{417}.

To conclude we have found that $\vb s|s\hb$ is only well defined and finite
when
\be
&&\frac{e}{f}=\frac{d}{c}=-\frac{b}{a}=-ir
\e{423}
where $r$ is a real constant which is finite and different from zero.
In this case our
complex operators have the form \be
&&\phi_a\equiv a(p_a+ir\pi_a),\;\;\;\xi^a\equiv
\frac{1}{2a^*}(ix^a-\frac{1}{r} v^a)\nn\\ &&\rho^a\equiv
\frac{1}{2a^*}(\ca^a+i\frac{1}{r}\bapet^a),\;\;\;k_a\equiv a(\pet_a+ir\baca_a)
\e{424}
Notice that if $r$ is imaginary the complex operators \r{424} are essentially
hermitian which is the reason why we found an undefined expression for
$\vb s|s\hb$ in this
case. ( $|s\hb$ is then not a well defined inner product state but
rather a state like $|\phi\hb$ in sections 2 and 3.)

That either $x^a$ or $ v^a$ should be chosen to have  imaginary eigenvalues was
one of the basic quantization rules found in \cite{Proper}. The basic reason
for
this is that the complex bosonic operators $\phi_a$ and $\xi^a$ span a
Fock space with
half positive and half indefinite metric states (see section 5).

We end this section by constructing a representation of $|s\hb$ in the form
\r{31},
\ie
\be
&&|s\hb=e^{[Q, \psi]}|\phi\hb
\e{425}
where $\psi$ is a gauge fixing fermion of the  form \r{32},
 and where $|\phi\hb$ satisfies one of the conditions in \r{26} for
$|\phi\hb_1$
or $|\phi\hb_2$.
We notice then that for $\al\beta>0$ \r{414} implies
\be
&&\phi'_a|\phi\hb=\rho^{'a}|\phi\hb=k'_a|\phi\hb=\xi^{'a}|\phi\hb=0
\e{427}
where
\be
&\phi'_a\equiv e^{-[Q, \psi]}\phi_a e^{[Q, \psi]}=&
ap_a(\cos{\sqrt{\al\beta}}-\al
r\frac{\sin{\sqrt{\al\beta}}}{\sqrt{\al\beta}})+\nn\\&&+ira
\pi_a(\cos{\sqrt{\al\beta}}+
\frac{\beta}{r}
\frac{\sin{\sqrt{\al\beta}}}{\sqrt{\al\beta}})\nn\\&\xi^{'a}\equiv e^{-[Q,
\psi]}\xi^a e^{[Q, \psi]}=&\frac{1}{2a^*}ix^a(\cos{\sqrt{\al\beta}}+
\frac{\beta}{r}
\frac{\sin{\sqrt{\al\beta}}}{\sqrt{\al\beta}})-\nn\\&&-
\frac{1}{2ra^*} v^a(\cos{\sqrt{\al\beta}}-\al
r\frac{\sin{\sqrt{\al\beta}}}{\sqrt{\al\beta}})\nn\\
&\rho^{'a}\equiv e^{-[Q, \psi]}\rho^a e^{[Q, \psi]}=&
\frac{\ca^a}{2a^*}(\cos{\sqrt{\al\beta}}+\frac{\beta}{r}
\frac{\sin{\sqrt{\al\beta}}}{\sqrt{\al\beta}})+\nn\\&&+
i\frac{1}{2ra^*}\bapet^a(\cos{\sqrt{\al\beta}}-\al r
\frac{\sin{\sqrt{\al\beta}}}{\sqrt{\al\beta}})\nn\\&k'_a\equiv e^{-[Q,
\psi]}k_a
e^{[Q, \psi]}=&a\pet_a(\cos{\sqrt{\al\beta}}-\al r
\frac{\sin{\sqrt{\al\beta}}}{\sqrt{\al\beta}})+\nn\\&&+
ira\baca_a(\cos{\sqrt{\al\beta}}+\frac{\beta}{r}
\frac{\sin{\sqrt{\al\beta}}}{\sqrt{\al\beta}})
\e{428}
Hence, if $|\phi\hb$ satisfies the condition for $|\phi\hb_1$ in \r{26} then
\r{428}
requires
\be
&&r=\frac{\sqrt{\al\beta}}{\al \tan{\sqrt{\al\beta}}}
\e{429}
If on the other hand $|\phi\hb$ satisfies the condition for $|\phi\hb_2$ in
\r{26}
then \r{428} requires \be
&&r=-\frac{\beta \tan{\sqrt{\al\beta}}}{\sqrt{\al\beta}}
\e{430}
For $\al\beta<0$ we have to make use of the replacement \r{37} in \r{428}.
We find then
correspondingly
\be
&&r=\frac{\sqrt{-\al\beta}}{\al \tanh{\sqrt{-\al\beta}}}
\e{431}
and
\be
&&r=-\frac{\beta \tanh{\sqrt{-\al\beta}}}{\sqrt{-\al\beta}}
\e{432}

Since $r=1/\al'$ for $|\phi\hb_1$ and $r=-\beta'$ for $|\phi\hb_2$ where $\al'$
and
$\beta'$ are given by \r{317} or \r{318} we notice that a finite nonzero
$r$ in \r{424}
exactly excludes those values of $\al$ and $\beta$ for which the representation
\r{425} is not a singlet state and not an inner product state, \ie the
conditions
found in section 3. Notice also that to every consistent choice of $\al$ and
$\beta$
there is a corresponding $r$ and vice versa.
However, there does not exist any choice
of $\al$ and $\beta$ for a given $r$ for which \r{414} or \r{415} in the
representation \r{424} allows for {\em both} choices of "boundary" conditions
of
$|\phi\hb$ in \r{26}.

Finally, one may notice that the expressions \r{35} or \r{36} are essentially
obtained when \r{429} or \r{430} is inserted into \r{424}. In particular yields
$r=1/\al$ and $r=-\beta$ in \r{424} essentially the two sets in \r{29}.

 \setcounter{equation}{0}
\section{The coBRST charge for the simple abelian model}
In order to construct a general coBRST charge operator for the simple abelian
model of section 2  using the definition \r{14}  we   need to
construct the metric operator $\eta$ in \r{15}. This in turn requires us  to
diagonalize the oscillators $\phi,\;\xi,\;\rho$ and $k$ in \r{424}.
Starting from a general linear ansatz we find  the following expressions
 for diagonalized oscillators (suppressing indices)
\be
&&a=R^{-1}(\xi+M\phi),\;\;\;b=UR^{-1}(\xi-M^{\dag}\phi)\nn\\
&&A=S^{-1}(\rho+N k),\;\;\;B=VS^{-1}(\rho-N\dagg k)
\e{501}
They satisfy the commutator algebra (the non-zero part)
\be
&&[a_a, a_b^{\dag}]_-=\del_{ab},\;\;\;[b_a,
b_b^{\dag}]_-=-\del_{ab},\;\;\;[a_a,
b_b^{\dag}]_-=0\nn\\
&&[A_a, A_b^{\dag}]_+=\del_{ab},\;\;\;[B_a,
B_b^{\dag}]_+=-\del_{ab},\;\;\;[A_a,
B_b^{\dag}]_+=0
\e{502}
In \r{501} the vector operators $a,\;b,\;A,\;B,\;\phi$ and $k$ have lower
indices
while $\xi$ and $\rho$ have upper ones. $U$ and $V$ are arbitrary unitary
matrices
and $R,\;S,\;M$ and $N$ are arbitrary complex invertible matrices. However, the
hermitian parts of $M$ and $N$ are determined by $R$ and $S$ through the
relations
\be
&&M+M^{\dag}=RR\dagg,\;\;\;N+N^{\dag}=SS\dagg
\e{503}
Hence, the
hermitian parts of $M$ and $N$ are strictly positive.
The oscillators in \r{501} are obviously noncovariant in general (except when
the metric that raises and lowers indices is euclidean
\ie $g_{ab}=\pm\del_{ab}$).

The metric operator $\eta$ has now the form
\be
&&\eta=\eta_B\eta_F
\e{504}
where \cite{Gen}
\be
&&\eta_B=\exp{(i\pi\sum_{a=1}^{n}b_a^{\dag}b_a)},
\;\;\;\eta_F=\prod_{a=1}^{n}(1+2B_a^{\dag}B_a)
\e{505}
which imply
\be
&&\eta b_a\eta=-b_a,\;\;\;\eta B_a\eta=-B_a
\e{506}
For the original oscillators $\xi,\;\phi,\;\rho$ and $k$ this implies using
\r{501}
(notice that $M\dagg(M+M\dagg)^{-1}M=M(M+M\dagg)^{-1}M\dagg$ and
$N\dagg(N+N\dagg)^{-1}N=N(N+N\dagg)^{-1}N\dagg$)
\be
&&\eta\xi\eta=(M\dagg-M)(M+M\dagg)^{-1}\xi+
2M(M+M\dagg)^{-1}M\dagg\phi\nn\\
&&\eta\phi\eta=2(M+M\dagg)^{-1}\xi+ (M+M\dagg)^{-1}(M-M\dagg)\phi\nn\\
&&\eta\rho\eta=(N\dagg-N)(N+N\dagg)^{-1}\rho+
2N(N+N\dagg)^{-1}N\dagg k\nn\\
&&\eta k\eta=2(N+N\dagg)^{-1}\rho+
(N+N\dagg)^{-1}(N-N\dagg)k
\e{507}
Remarkably enough these expressions do not involve the matrices $R,\;S,\;U$ and
$V$ in \r{501}.  All arbitrariness lies in the matrices $M$ and $N$ which
partly are
determined by $R$ and $S$ through the relations \r{503}.

Formula \r{14} yields now  the general coBRST charge operator
\be
^*Q&\equiv&\eta Q\eta=\eta(\rho^{a\dag}\phi_a + \phi_a^{\dag}\rho^a)\eta=\nn\\
&=&4k\dagg N(N+N\dagg)^{-1}N\dagg(M+M\dagg)^{-1}\xi+\nn\\
&&+4\xi\dagg(M+M\dagg)^{-1}N(N+N\dagg)^{-1}N\dagg
k+\nn\\
&&+\rho\dagg(N+N\dagg)^{-1}(N-N\dagg)(M+M\dagg)^{-1}(M-M\dagg)\phi+ \nn\\
&&+\phi\dagg(M\dagg-M)(M+M\dagg)^{-1}(N\dagg-N)(N+N\dagg)^{-1}\rho+\nn\\
&&+2\rho\dagg(N+N\dagg)^{-1}(N-N\dagg)(M+M\dagg)^{-1}\xi+\nn\\
&&+2\xi\dagg(M+M\dagg)^{-1}(N\dagg-N)(N+N\dagg)^{-1}\rho+\nn\\
&&+2k\dagg N(N+N\dagg)^{-1}N\dagg(M+M\dagg)^{-1}(M-M\dagg)\phi+\nn\\
&&+2\phi\dagg(M\dagg-M)(M+M\dagg)^{-1}N(N+N\dagg)^{-1}N\dagg
k
\e{508}
This expression satisfies
\be
\triangle&\equiv&[Q,\,^*Q]_+=[Q,k\dagg N'(M')^{-1}\xi+
\xi\dagg(M')^{-1}N'
k]=\nn\\
&=&\phi\dagg N'(M')^{-1}\xi+k\dagg
N'(M')^{-1}\rho+\rho\dagg(M')^{-1}N'
k+\xi\dagg(M')^{-1}N'
\phi=\nn\\
&=&\xi N'(M')^{-1}\phi\dagg-\rho N'(M')^{-1}k\dagg-k(M')^{-1} N'
\rho\dagg+\phi(M')^{-1} N'
\xi\dagg \e{509}
where
\be
&&M'\equiv \halv\left(M+M\dagg\right),\;\;\;N'\equiv
2N(N+N\dagg)^{-1}N\dagg \e{5091}
The properties of the matrices $M$ and $N$ require $M'$ and
$N'$ to be invertible.
As a consequence $\triangle|s\hb=0$
imply \r{414} or \r{415}, which is  also a consequence of $Q|s\hb=\,^*Q|s\hb=0$
(see
\eg \cite{KvH}). The original state space, $V$, is spanned by eigenstates to
$\triangle$ with positive integers as eigenvalues. (This is at least true when
$M'$
and $N'$ commute.) This leads to the Hodge decomposition \r{16} (see
\eg \cite{KvH}).

Since only the first two terms in
\r{508} contribute to the commutator \r{509},  it is natural, and always
allowed, to
choose the matrices $M$ and $N$ to be hermitian, in which case \r{508}
reduces to
\be
&&^*Q=k\dagg NM^{-1}\xi+
\xi\dagg M^{-1}N
k
\e{510}
and \r{507} becomes
\be
&&\eta\xi\eta=M\phi,\;\;\;
\eta\phi\eta=M^{-1}\xi,\;\;\;
\eta\rho\eta=N k,\;\;\;
\eta k\eta=N^{-1}\rho
\e{511}
If we furthermore choose $N=\la M$ where $\la$
is a real positive constant then $^*Q$
acquires the covariant form
\be
&&^*Q=\la \left(k_a\dagg\xi^a+
\xi^{\dag a}k_a\right)
\e{512}
Such expressions for coBRST are given in the literature (see \eg
ref.\cite{RRy}).

By means of \r{424} the expression \r{512} may be rewritten in terms of the
original
variables. We find then
\be
&&^*Q=\la \left(r\baca_ax^a-\frac{1}{r}\pet_a v^a\right)
\e{513}
where $r$ is the real constant in \r{424}. Such an expression for the coBRST
charge seems not to have been given before.   It
differs \eg from the suggestion given in \cite{Solv}.

The expression \r{513} when compared with \r{32} shows that the coBRST charge
in
this case may be viewed as a fermionic gauge fixing variable. This is very
natural
since both $\psi$ and $^*Q$ have to do with gauge fixing. In fact, if $\psi$
in the
representation \r{31} is chosen to be the coBRST charge \r{513} we find the
expressions \r{316} with
\be
\al'=-\frac{1}{r}\tanh\la,\;\;\;\beta'={r}\tanh\la
\e{5131}
Thus, $\al'$ and $\beta'$ are then finite and non-zero for any finite and
non-zero
$r$ and $\la$ in \r{513}, which means that \r{513} is always a good gauge
fixing
fermion.

It is natural to expect that also  the more general expressions \r{508} or
\r{510}
should be possible to use as a gauge fixing fermion. In terms of the original
variables \r{510} becomes
\be &&^*Q=r\baca_a T^{ab}x^b-\frac{1}{r}\pet_a T^{ab}
v^b+\baca_a L^{ab} v^b+\pet_a L^{ab}x^b \e{514}
 where $T^{ab},\;L^{ab}$  are real
matrices  and $T^{ab}$ is invertible.
(If $NM^{-1}$ in \r{510} is real then $L=0$ and
$T=NM^{-1}$.) Notice that the parameter $r$ is related to the choice of
oscillator
basis and the matrices $T^{ab},\;L^{ab}$ to the choice of diagonal
representation of
this oscillator basis. $r$ is more important since it is related to the
choice of
vacuum. Anyway, if \r{514} is chosen to be a gauge fixing fermion, then
we find
\be
[Q,\,^*Q]\equiv A(T)+B(L)
\e{515}
where
\be
&&A(T)\equiv r\pi_aT^{ab}x^b-\frac{1}{r}p_a T^{ab} v^b+ir\baca_a
T^{ab}\ca^b-i\frac{1}{r}\pet_aT^{ab}\bapet^b   ,\nn\\
&&B(L)\equiv\pi_aL^{ab}  v^b+p_a
L^{ab}x^b +i\baca_aL^{ab} \bapet^b+i\pet_aL^{ab} \ca^b \e{5151}
We notice now that $B(L)$ satisfies
\be
B(L)|\phi\hb_1=B(L)|\phi\hb_2=0
\e{516}
and that
\be
[A(T),B(L)]_+=0
\e{517}
if the matrices
$T^{ab}$ and $L^{ab}$ commute. In this latter case we find therefore
\be
e^{[Q,
\,^*Q]}|\phi\hb_l=e^{A(T)}|\phi\hb_l\equiv e^{[Q, \psi]}|\phi\hb_l
\e{518}
 where
\be
\psi\equiv r\baca_a T^{ab}x^b-\frac{1}{r}\pet_aT^{ab} v^b
\e{519}
 which should be an allowed form
in general since it is a linear combination of \r{26}. (The condition \r{517}
on the
matrices $T$ and $L$ can probably be weakened.)

\setcounter{equation}{0}
\section{Generalizations to nonabelian theories}
So far we have only performed a detailed analysis of simple abelian models.
In order
to demonstrate that our results are not special properties of
such models in this section
we shall   consider the class of nonabelian models in which the gauge
group is a general Lie group.  Within the
 corresponding BRST invariant models the standard BFV-BRST charge
is  given by ($a, b, c =1,\ldots,n<\infty$)\cite{BV}
\be
&&Q=\theta_a\ca^a-\frac{1}{2}iU_{bc}^{\;\;\;a}\pet_a
\ca^b\ca^c-\frac{1}{2}iU_{ab}^{\;\;\;b}\ca^a + \bapet_a\pi^a
\e{601}
where $\theta_a$ are the hermitian bosonic gauge generators
(constraints) satisfying
\be
&[\theta_a, \theta_b]_{-}=iU_{ab}^{\;\;\;c}\theta_c
\e{602}
where $U_{ab}^{\;\;\;c}$ are the real structure constants.
In direct analogy with what
we had for the abelian models we propose now the following
singlet representation for
models invariant under \r{601}
\be
&&|s\hb_l=e^{[Q, \psi]}|\phi\hb_l,\;\;\;l=1,2
\e{603}
where the  gauge fixing fermion $\psi$  is given by
 \be
&&\psi=\al\pet_a v^a+\beta\baca_a x^a
\e{604}
where in turn the gauge fixing variables $x^a$ are chosen such that they
commute with
all variables except the constraints $\theta_a$ and satisfy  the conditions \be
&&|x^a, x^b]=0,\;\;\;[x^a, \theta_b]=iM^a_{\;\;b}
\e{605}
where $M^a_{\;\;b}$ is an invertible matrix operator.
$|\phi\hb_1$
and $|\phi\hb_2$ satisfy here
\be
&&x^a|\phi\hb_1=\ca^a|\phi\hb_1=\baca_a|\phi\hb_1=\pi_a|\phi\hb_1=0\nn\\
&& v^a|\phi\hb_2=\bapet^a|\phi\hb_2=\pet_a|\phi\hb_2=\left(\theta_a+i\halv
U_{ab}^{\;\;\;b}\right)
 |\phi\hb_2=0 \e{606}
Notice that only the last conditions on $|\phi\hb_2$ differ from \r{26}.
($U_{ab}^{\;\;\;b}=0$ for
unimodular gauge groups as in \eg  Yang-Mills.) The new conditions follow
from the fact that \be
&&[Q, \pet_a]=\theta_a+\theta_a^{gh},\;\;\;\theta_a^{gh}\equiv
i\halv U_{ab}^{\;\;\;c}\left(\pet_c\ca^b-\ca^b\pet_c\right)
 \e{607}
and
\be
&&[Q, \pet_a]|\phi\hb_2=\left(\theta_a+i\halv U_{ab}^{\;\;\;b}
\right)|\phi\hb_2
\e{608}

As in section 3 we write now \r{603} in the following form
\be
&&|s\hb_l=e^{\al K_1+\beta K_2}|\phi\hb_l,\;\;\;l=1,2
\e{609}
 where now the hermitian operators $K_1$ and $K_2$ are given by
\be
&&K_1\equiv[Q, \pet_a v^a]= (\theta_a+\theta_a^{gh})v^a+i\pet_a\bapet^a\nn\\
&&K_2\equiv[Q, \baca_a x^a]=x^a \pi_a+i\baca_a \ca^b M^a_{\;\;b}
\e{610}
(cf \r{312}).
	In addition, we introduce the hermitian operator $K_3$ defined by
\be
&&K_3\equiv i\halv[K_1,
K_2]_-=\halv\left( \pi_a
M^a_{\;\;b}v^b-(\theta_a+\theta_a^{gh})x^a\right)+\nn\\
&&+\halv\left(i\baca_a M^a_{\;\;b}\bapet_b-i\pet_a
M^a_{\;\;b}\ca^b  -\baca_a\ca^b v^c[\theta_b,
M^a_{\;\;c}]   \right)
 \e{611}
In obtaining the last equality we have made use of the Jacoby identities
\be
&&U_{cb}^{\;\;\;d} M^a_{\;\;d}=i[\theta_b, M^a_{\;\;c}]-i[\theta_c,
M^a_{\;\;b}]
\e{612}
We notice the properties
\be
&&K_2|\phi\hb_1=K_1  |\phi\hb_2=K_3    |\phi\hb_1=K_3|\phi\hb_2=0
\e{613}
which are identical to \r{315} which we had  in the abelian case.  However, in
distinction to what we had there
the $K_i$ operators \r{610}-\r{611} do not satisfy a   closed commutator
algebra. On
the other hand, in appendix C it is shown  that
 \be
&&[K_2, K_3]=-iK_2,\;\;\;[K_1, K_3]|\phi\hb_1=iK_1|\phi\hb_1,
\e{614}
provided $x^a$ are chosen to be canonical coordinates on the group manifold.
($M^a_{\;\;b}$ depends then only on $x^a$.) In fact, in this case $K_i$ satisfy
effectively an $SL(2,R)$ algebra on $|\phi\hb_l$ and it is straight-forward to
derive
the  relations (see appendix C)
\be
|s\hb_1=e^{\al K_1+\beta K_2}|\phi\hb_1=e^{\al'
K_1}|\phi\hb_1,\;\;\; |s\hb_2=e^{\al K_1+\beta K_2}|\phi\hb_2=
e^{\beta' K_2}|\phi\hb_2
\e{615}
where
\be
&&\al'=\al\,\frac{\tan{\sqrt{\al\beta}}}{\sqrt{\al\beta}},
\;\;\;\beta'=\beta\,\frac{\tan{\sqrt{\al\beta}}}{\sqrt{\al\beta}}
\e{616}
for $\al\beta>0$ and
\be
&&\al'=\al\,\frac{\tanh{\sqrt{-\al\beta}}}{\sqrt{-\al\beta}},
\;\;\;\beta'=\beta\,\frac{\tanh{\sqrt{-\al\beta}}}{\sqrt{-\al\beta}}
\e{617}
for $\al\beta<0$. Eqs.\r{615}-\r{617} are identical to
\r{316}-\r{318} in the abelian
case! The conditions for $|s\hb_1$ and
$|s\hb_2$ to be inner product
states should, therefore, be identical to the ones obtained in section 3
\ie $\al'\neq0$ and $\beta'\neq0$
respectively. Notice, however, that there exists no {\em general} proof that
$|s\hb_1=e^{\al' K_1}|\phi\hb_1$ and $|s\hb_2= e^{\beta' K_2}|\phi\hb_2$ are
inner
product states in the nonabelian case although this is  expected  to be the
case
 (see refs.\cite{Proper,Gauge,Quaade} and below).

{}From the conditions \r{606} on $|\phi\hb_l$ we may derive the conditions
satisfied by
$|s\hb_1$ and $|s\hb_2$  corresponding to \r{414} in
the abelian case. If we restrict ourselves to the case when  $x^a$ are
canonical
coordinates on the group manifold then eq.\r{615} is valid and we have
 \be
&&x^{'a}|s\hb_1=\ca^{'a}|s\hb_1=\baca'_a|s\hb_1=\pi'_a|s\hb_1=0\nn\\
&& v^{''a}|s\hb_2=\bapet^{''a}|s\hb_2=\pet''_a|s\hb_2=\left(\theta''_a+i\halv
U_{ab}^{\;\;\;b}\right)
 |s\hb_2=0
\e{618}
where
\be
&x^{'a}\equiv e^{\al'K_1}x^a
e^{-\al'K_1}=&e^{\al'\theta_av^a}x^ae^{-\al'\theta_av^a}\equiv
f^a(x,-i\al'v)\nn\\&
\ca^{'a}\equiv e^{\al'K_1}\ca^a
e^{-\al'K_1}=&A^a_{\;\;b}(-i\al'v)\ca^b-i\al'
(M^{-1})^a_{\;\;b}(-i\al'v)\bapet^b\nn\\&
  \baca'_a\equiv e^{\al'K_1}\baca_a
e^{-\al'K_1}=&\baca_a+i\al'(M^{-1})^a_{\;\;b}(i\al'v)\pet_b\nn\\
&\pi'_a\equiv e^{\al'K_1}\pi_a
e^{-\al'K_1}=&\pi_a+i\al'(M^{-1})^b_{\;\;a}(i\al'v)
\left(\theta_a+\theta_a^{gh}\right)
-\nn\\&&-\al'\frac{\partial}{\partial v^d}(M^{-1})^c_{\;\;a}
(i\al'v)\pet_c\bapet^d
\e{619}
\be
&&v^{''a}\equiv e^{\beta'K_2}v^a e^{-\beta'K_2}=v^a-i\beta'x^a\nn\\
&&
\bapet^{''a}\equiv e^{\beta'K_2}\bapet^a
e^{-\beta'K_2}=\bapet^a-i\beta'\ca^bM^a_{\;\;b}(x)\nn\\
 &&\pet''_a\equiv
e^{\beta'K_2}\pet_a e^{-\beta'K_2}=\pet_a+i\beta'\baca_bM^b_{\;\;a}(x)\nn\\
&&
\theta''_a\equiv e^{\beta'K_2}\theta_a
e^{-\beta'K_2}=\theta_a+i\beta'\baca_b\ca^c[M^b_{\;\;c}(x), \theta_a]
 \e{620}
Notice that $f^a(x,-i\al'v)$ are also canonical coordinates on
the group manifold obtained by
two successive transformations, one with coordinates $x^a$ and one with
$-i\al'v^a$.
$A^a_{\;\;b}(-i\al'v)=(M^{-1})^a_{\;\;c}(i\al'v)M^c_{\;\;b}(-i\al'v)$
is the adjoint matrix representation of the group. Some properties of the
matrices $M^a_{\;\;b}$ are given in appendix C.
In deriving \r{619} we have made use of
the relation
\be
&&e^{\al'K_1}=e^{i\al'(M^{-1})^a_{\;\;b}(i\al'v)\pet_a\bapet^b}
e^{\al'\left(\theta_a+\theta_a^{gh}\right)v^a}
\e{6201}
which may be obtained from ref.\cite{Quaade}. The expressions for ${\ca'}^a$
and
$\pi'_a$ were also obtained in \cite{Simple} (formulas (4.5) and (4.8)).
One may
notice  that
 \r{619} and \r{620} are  nonlinear in distinction to the linear
expressions \r{29} in the abelian case.  From \r{619} and \r{620}
it follows now that
 \be
&&[x^{'a},
(\pi'_b)^{\dag}]=\al'\left\{\frac{\partial
f^a(x,-i\al'v)}{\partial(-i\al'v)^b}+M^d_{\;\;c}(x)(M^{-1})^c_{\;\;b}(-i\al'v)
\frac{\partial
f^a(x,-i\al'v)}{\partial x^d}\right\}
\nn\\&& [\ca^{'a},
(\baca'_b)^{\dag}]=-i\al'(M^{-1})^c_{\;\;b}(-i\al'v)\left(\del^a_b+
A^a_{\;\;c}(-i\al'v)\right)
\e{621}  and
\be
&&[v^{''a},
(\theta''_b)^{\dag}]=\beta'M^a_{\;\;b}(x)
,\;\;\;
[\bapet^{''a},
(\pet''_b)^{\dag}]=-2i\beta'M^a_{\;\;b}(x)
 \e{622}
Since one may easily convince oneself that the matrices on the right-hand sides
of
\r{621} and \r{622} are invertible the conditions for $|s\hb_1$ and $|s\hb_2$
to
be
singlet states are   $\al'\neq0$ and $\beta'\neq0$ respectively. These
conditions
are exactly the same as the ones  we had in the abelian case, as well as those
which were required for $|s\hb_1$ and $|s\hb_2$ to be inner product states.

We are now in principle able to calculate the coBRST charge in the same way as
we did
for the abelian models. However, since this requires us to diagonalize the
"oscillators" \r{619} and \r{620} which is quite cumbersome,  we shall not
do that
here (see below, however). Instead we shall just demonstrate that a coBRST
charge  of
the same form as we had in the abelian case \ie \be &&^*Q=\lambda\left(
r\baca_ax^a-\frac{1}{r}\pet_av^a\right)
\e{623}
will leave the singlet states \r{603} invariant under the expected conditions.
(Below it will be proved that \r{623} actually is an appropriate coBRST
charge.)  To this end let us define $^*Q'$ and $^*Q''$ by \be
&&^*Q'\equiv e^{-\al'K_1}\,^*Qe^{\al'K_1}=\nn\\&&
=\lambda\left(
r\left(\baca_a-i\al'(M^{-1})^b_{\;\;a}(-i\al'v)\pet_b\right)f^a(x,i\al'v)-
\frac{1}{r}\pet_av^a\right)\nn\\&&^*Q''\equiv
e^{-\beta'K_2}\,^*Qe^{\beta'K_2}=\nn\\&&=\lambda\left(
r\baca_ax^a-\frac{1}{r}\left(\pet_av^a
+i\beta'\pet_ax^a-i\beta'\baca_bM^b_{\;\;a}(x)v^a+{\beta'}^2\baca_b
M^b_{\;\;a}(x)x^a\right)\right)
\e{624}
We have then by means of the properties
$M^a_{\;\;b}(0)=\del^a_b$, $f^a(0,y)=y^a$,
and \r{606}
 \be
&&^*Q|s\hb_1=e^{\al'K_1}\,^*Q'|\phi\hb_1=e^{\al'K_1}\lambda\left(
r{\al'}^2\pet_av^a-
\frac{1}{r}\pet_av^a\right)|\phi\hb_1\nn\\
&&^*Q|s\hb_2=e^{\beta'K_2}\,^*Q''|\phi\hb_2=e^{\beta'K_2}\lambda\left(
r\baca_ax^a-\frac{1}{r}{\beta'}^2\baca_a
x^a\right)|\phi\hb_2
\e{625}
Hence,  we have
\be
&&^*Q|s\hb_1=\,^*Q|s\hb_2=0
\e{626}
provided
$r=\pm1/{\al'}$ and $r=\pm\beta'$ respectively.

Remarkably enough there exists a simple abelianization of the BRST charge
\r{601} which allows us to make use of all results of our analysis of abelian
models
also for the nonabelian models considered here. This abelianization is
performed by
means of $x^a$ as canonical coordinates  on the group manifold and the
matrix $M^a_{\;\;b}(x)$ as follows:
  According to \r{c5} in appendix C we may define  hermitian
conjugate momenta to $x^a$ by
\be
&&p_a=(M^{-1})^b_{\;\;a}(x)\theta_b+i\halv
(M^{-1})^b_{\;\;a}(x)\partial_cM^c_{\;\;b}(x)
\e{627}
We have then
\be
&&\theta_a=\halv\left(p_bM^b_{\;\;a}(x) + M^b_{\;\;a}(x)p_b  \right)
\e{628}
Consider furthermore a unitary transformation which only affects
$\eta^a$, $\pet_a$,
and $p_a$, and which is of the following form
\be
&&\tca^a=M^a_{\;\;b}(x)\ca^b,\;\;\;\tpet_a=(M^{-1})^b_{\;\;a}(x)\pet_b\nn\\
&&\tpe_a=p_a+i\halv
\partial_aM^b_{\;\;c}(x)(M^{-1})^d_{\;\;b}(\ca^c\pet_d-\pet_d\ca^c)
 \e{629}
If one  inserts \r{628} into \r{601} and replaces $\ca^a$, $\pet_a$, and $p_a$
by
$\tca^a$, $\tpet_a$, and $\tpe_a$ using \r{629} then one finds
\be
&&Q=\tca^a\tpe_a+\pi_a\bapet^a
\e{630}
which is the BRST charge \r{21} for an abelian model. (A similar
abelianization of
classical Yang-Mills was considered in \cite{STH}.) In this way we may
now apply all
our results obtained for abelian models to the general nonabelian model
\r{601}. We
have, thus,  the representation \r{31} for the singlet states, \ie \be
&&|s\hb_l=e^{[Q, \psi]}|\phi\hb_l,\;\;\;l=1,2
\e{631}
where the  gauge fixing fermion $\psi$  is given by
 \be
&&\psi=\al\tpet_a v^a+\beta\baca_a x^a=\al
(M^{-1})^b_{\;\;a}(x)\pet_b v^a+\beta\baca_a x^a
\e{632}
and where $|\phi\hb_l$ satisfies \r{26} \ie
\be
&&x^a|\phi\hb_1=\tca^a|\phi\hb_1=\baca_a|\phi\hb_1=\pi_a|\phi\hb_1=0\nn\\
&& v^a|\phi\hb_2=\bapet^a|\phi\hb_2=\tpet_a|\phi\hb_2=\tpe_a
 |\phi\hb_2=0 \e{633}
Since $M^a_{\;\;b}$ is an invertible matrix operator one may easily show that
the
conditions \r{633} are equivalent to \r{606}. From our analysis of abelian
models we
have now that if $K_2$ is defined by \r{610} and
\be
&&K_1\equiv[Q, \pet_aM^a_{\;\;b}(x)v^b],\;\;\;K_3\equiv i\halv[K_1, K_2]
\e{634}
then $K_i$ will satisfy the $SL(2,R)$ algebra \r{314} exactly which was not
the case
above. The properties \r{613}, \r{615}-\r{617} are then easily verified.
This means that \r{631}
are singlet states under exactly the same conditions on $\al$ and
$\beta$ in \r{632} as
\r{603} are singlets for $\al$ and
$\beta$ in \r{604}. From section 5 we obtain the coBRST charge of the
general form \r{514}, \ie \be
 &&^*Q=\la\left(r\baca_a T^{a}_{\;\;b}x^b-\frac{1}{r}\tpet_a T^{a}_{\;\;b}
v^b\right)+\baca_a L^{a}_{\;\;b} v^b+\tpet_a L^{a}_{\;\;b}x^b
\e{635}
In particular with $T^{a}_{\;\;b}=M^{a}_{\;\;b}(x)$ and $L^{a}_{\;\;b}=0$
\r{635}
reduces exactly to \r{623} since
$M^{a}_{\;\;b}(x)x^b=x^a$.
Thus, we have showed that
 \r{623} is a coBRST charge. Notice that we equally well may choose
($T^{a}_{\;\;b}=\del^{a}_{b},\;L^{a}_{\;\;b}=0$)
 \be &&^*Q=\lambda\left(
r\baca_ax^a-\frac{1}{r}\pet_a(M^{-1})^a_{\;\;b}v^b\right)
\e{636}
In fact, the states \r{631} are invariant under \r{636} if $r=\pm1/{\al'}$ and
$r=\pm\beta'$ respectively, where $\al'$ and $\beta'$ are given by
\r{616} and \r{617}
where $\al$ and $\beta$ now are those in \r{632}.

\setcounter{equation}{0}
\section{Summary and conclusions}
In this paper we have considered gauge fixing and coBRST invariance
of both abelian and
nonabelian gauge theories. The  gauge theories were given in standard BFV-form
and
quantized on a state space $V$ with a nondegenerate inner product $\vb u|v\hb$.
This inner product of  $V$ was required to be a linear form on
a Hilbert space which means that $V$ is a Krein space \cite{Bog,RRy}.
This is a property
which always allows us
 to define a coBRST charge. The metric operator $\eta$ that relates $V$ with a
Hilbert space is expressed in terms of the indefinite oscillators in the theory
and
has the property $\eta^2=\bett$. The  coBRST charge
$^*Q$ is defined in terms of $\eta$
and the nilpotent BRST charge $Q$  by $^*Q\equiv\eta Q\eta$. The BRST
singlets, $|s\hb$, the states that represent the true physical degrees of
freedom and
which constitute a representation of the BRST cohomology ($|s\hb\in
\mbox{Ker} Q/\mbox{Im} Q$) are determined by the conditions
\be
  &&Q|s\hb=\,^*Q|s\hb=0
\e{700}
or equivalently
\be
&&\triangle|s\hb=0,\;\;\;
\triangle\equiv[Q,\,^*Q].
\e{701}
 The questions we have tried to answer in this paper
are the following ones:  What is the general BFV-form of the coBRST
charge $^*Q$ and
what is the general form of the gauge fixing fermions $\psi$ in the
representations of
BRST singlets found in \cite{Solv}, \ie $|s\hb=e^{[Q, \psi]}|\phi\hb$
where $|\phi\hb$
is a simple BRST invariant state? The answers to these two questions turned
out to be
interrelated since we have found that $\psi$ may be chosen to be equal to a
coBRST charge. Below we summarize our results and discuss their implications.

For the abelian models introduced in section 2 ($Q=\ca^ap_a+\bapet^a\pi_a$)
we found in
section 4 that the singlet states are determined by\be
&&\phi_a|s\hb=\rho^a|s\hb=k_a|s\hb=\xi^a|s\hb=0
\e{702}
or
\be
&&\phi^{\dag}_a|s\hb=\rho^{a\dag}|s\hb=k^{\dag}_a|s\hb=\xi^{a\dag}|s\hb=0
\e{703}
where
\be
&&\phi_a\equiv \frac{1}{\sqrt{2}}(p_a+ir\pi_a),\;\;\;\xi^a\equiv
\frac{1}{\sqrt{2}}(ix^a-\frac{1}{r} v^a)\nn\\ &&\rho^a\equiv
\frac{1}{\sqrt{2}}(\ca^a+i\frac{1}{r}\bapet^a),\;\;\;k_a\equiv
\frac{1}{\sqrt{2}}(\pet_a+ir\baca_a)
\e{704}
where in turn $r$ is a real constant different from zero.
Notice that the solutions of
\r{702} and \r{703} constitute two different representations. Which one is
realized
depends on the choice of the original state space $V$. A given $V$ will
only allow for
solutions of one of these conditions. One may notice that the two
solutions correspond
to solutions of \r{702} for   opposite signs of $r$ in \r{704}. Solutions
of \r{702}
for different $r$'s but with the same signs  are unitarily equivalent. We have
$|s\hb'_l=U(\ga)|s\hb_l$ where $\ga$ is a real constant and  where
$U(\ga)=U_1(\ga)U_4(\ga)$ or $U(\ga)=U_2\dagg(\ga)U_3(\ga)$ where in turn
the unitary
operators $U_1,\;U_2,\;U_3,\;U_4$ are defined in appendix B. $|s\hb'$
satisfies then
the same conditions as $|s\hb$ with $r$ replaced by $re^\ga$.

In section 5 we determined the general form of the coBRST charge for the
abelian
models of section 2.
The metric operator $\eta$ was then   expressed in terms of the
indefinite oscillators
in the theory which were identified by a diagonalization of the oscillators
in \r{704}.
  We
found then that $^*Q$ is not uniquely defined since $\eta$ may be defined in
several
different ways even for one given $r$ simply since the diagonalization of
\r{704} is
not unique. $^*Q$ for different signs of $r$'s are  related by
$^*Q\,\ra\:-^*Q$, and
$^*Q$ for different $r$'s but with the same signs  are related by a unitary
transformation of the form mentioned above. A simple form of $^*Q$ in terms
of the
original variables given in
 section 2 was found to be \be
&&^*Q=\la \left(r\baca_ax^a-\frac{1}{r}\pet_a v^a\right)
\e{7041}
where $\la$ is a real positive constant.

For the nonabelian models treated in section 6 we found essentially the same
results.
It is remarkable that although the oscillators in \r{704} then are nonlinear
in the
original variables the coBRST charge may still be of the same form as for
abelian
models. The general forms of coBRST found in section 4 suggests that the
general BFV
form of the coBRST charge is
\be
&&^*Q=\baca_a\chi^a-\pet_a\Lambda^a
\e{7042}
where $\chi^a$ and $\Lambda^a$ are gauge fixing conditions to the gauge
generators
and the conjugate momenta to the Lagrange multipliers respectively. In
fact, they must
be related to the natural gauge fixing variables $x^a$ and the Lagrange
 multipliers
$v^a$ by positive matrices. However, since the coBRST charge is nilpotent
the form
\r{7042} requires the gauge conditions $\chi^a$ and $\Lambda^a$ to be
abelian. The
most general nilpotent coBRST charge will  allow for gauge conditions
which are
in involution. However,  in this case there are nonlinear terms in the
ghosts on the
right-hand side of \r{7042} (cf the construction of a nilpotent BRST charge
\cite{BF1,BF2}).

We have  investigated the
properties of the representations
\be
&&|s\hb_l=e^{[Q, \psi]}|\phi\hb_l,\;\;\;l=1,2
\e{705}
for the  singlet states found in \cite{Solv} in the case when
the gauge fixing fermion
$\psi$ has the form \be
&&\psi=\al\pet_a v^a+\beta\baca_a x^a
\e{706}
where $\al$ and $\beta$ are real constants,
and when $|\phi\hb_l$ is chosen to satisfy the conditions in \r{26} or \r{618}.
We have then found that (see appendices A and C)
\be
|s\hb_1=e^{\al' [Q, \pet_av^a]}|\phi\hb_1,\;\;\;
|s\hb_2=e^{\beta' [Q, \baca_ax^a]}|\phi\hb_2
\e{7061}
where
\be
&&\al'=\al\,\frac{\tan{\sqrt{\al\beta}}}{\sqrt{\al\beta}},
\;\;\;\beta'=\beta\,\frac{\tan{\sqrt{\al\beta}}}{\sqrt{\al\beta}}
\e{7062}
for $\al\beta>0$ and
\be
&&\al'=\al\,\frac{\tanh{\sqrt{-\al\beta}}}{\sqrt{-\al\beta}},
\;\;\;\beta'=\beta\,\frac{\tanh{\sqrt{-\al\beta}}}{\sqrt{-\al\beta}}
\e{7063}
for $\al\beta<0$. (The limit $\al\beta\,\ra\,0$ yields $\al'=\al$ and
$\beta'=\beta$.) For abelian models  it follows then from ref.\cite{Proper}
that
$|s\hb_1$ and $|s\hb_2$ are inner product states if $\al'$ and
$\beta'$ are finite and
non-zero which in turn requires $\al\neq0$ and $\beta\neq0$ respectively
together with $\sqrt{\al\beta}\neq n\pi/2$ for any positive integer $n$.
Exactly under these conditions $|s\hb_1$ and $|s\hb_2$  are also  singlet
states. In fact, $|s\hb_1$ and $|s\hb_2$ satisfy the singlet conditions
\r{702} if
$r=1/\al'$ and $r=-\beta'$ respectively. The results \r{7061} shows that
there are
many formally different representations \r{705} which really are equal (\ie
many
different $\al$ and $\beta$ in \r{706} lead to the same $\al'$ and $\beta'$ in
\r{7061}).
Notice that both $|s\hb_1$ and $|s\hb_2$ in \r{7061} cannot satisfy the singlet
conditions \r{702} for a given $r$ since this requires $r=1/\al'$ and
$r=-\beta'$ which implies  $\al'\beta'=-1$ which has no solution. However, both
$|s\hb_1$ and $|s\hb_2$ can be coBRST invariant under the same coBRST charge.
Invariance under \r{7041} for a given $r$ requires $\al\beta>0$ and
$\tan\al\beta=1$
\ie $r=\pm\sqrt{\beta/\al}$. (These conditions follow from the fact
that $r=\pm 1/\al'$
and $r=\pm \beta'$ allow for $\al'\beta'=1$.) Essentially the same results were
also
found for the nonabelian models in section 6.

There are certainly still more involved forms for the gauge fixing fermions
$\psi$
than \r{706} which are allowed in \r{705}. For the simple abelian
theory we could \eg
consider $\psi=\pet_aT^{ab} v^b+\baca_aS^{ab}x^b$ where $T^{ab}$ and $S^{ab}$
are
real, invertible matrices. The analysis of this case is
 much more involved than the one of \r{706}. One may notice that
such a $\psi$
is allowed for either $T^{ab}=0$ or $S^{ab}=0$. Furthermore,  if
$T^{ab}$ and $S^{ab}$
are symmetric and commuting one may prove that the
representation \r{705} is a singlet state up  similar conditions to the ones we
have for \r{706} using exactly the same analysis we have used for \r{706}.
An example
of such a gauge fixing is also considered for the nonabelian models in
section 6.
This suggests that even a gauge fixing fermion of the general BFV form (see \eg
\cite{BF1}), \ie $\psi=\baca_a\chi^a+\pet_a\Lambda^a$ in representations like
\r{705} do in fact yield singlet states.

One of the important results of our paper is that
 the coBRST charge is of the form of an allowed gauge fixing
fermion. We may therefore replace
 $\psi$ by a $^*Q$ in the representation
\r{705} in which case we have \be
&&|s\hb_l=e^{[Q, \,^*Q]}|\phi\hb_l\equiv e^{\triangle}|\phi\hb_l,\;\;\;l=1,2
\e{707}
Both for the abelian and nonabelian models our results show that
the choice \r{7041}
of the coBRST charge always makes \r{707}   a singlet state with a finite norm.
However, one may notice that this singlet state is not coBRST invariant under
the
same $^*Q$ since $|\phi\hb_l$ is never coBRST invariant by itself ($^*Q$
commutes with $\triangle$). On the other hand, $|s\hb_1$ ($|s\hb_2$) in \r{707}
is
coBRST invariant under a different coBRST charge, $^*Q'$,
obtained by the replacement
$r\,\ra\,\pm r/(\tanh \la)$ ($r\,\ra\,\pm r \tanh \la$) in $^*Q$.
Now $^*Q'$ and $^*Q$
are related by a unitary transformation involving a unitary
operator of the last form
in \r{a8}. This means that there are always  unitary operators
$U_{(l)}$ such that the
singlet states  \be &&|s\hb'_l\equiv U_{(l)}e^{[Q,
\,^*Q]}|\phi\hb_l,\;\;\;l=1,2
\e{7071}
 are invariant under $^*Q$. $U_{(1)}$ and $U_{(2)}$ may \eg be chosen to be
 $U\dagg_2(\ga)U_3(\ga)$ in appendix B with $\ga=\ln(\tanh\ga)$
and $\ga=-\ln(\tanh\ga)$ respectively.

A further intriguing feature of the representation \r{705} was discovered in
sections 3 and 6.   In the abelian case with \r{705}  written as (see
 \r{321}) \be &&|s\hb_l=e^{\al K_1+\beta K_2}|\phi\hb_l,\;\;\;l=1,2
\e{7072}
where $K_1\equiv[Q, \pet_av^a]$ and $K_2\equiv[Q, \baca_ax^a]$,
we found that $K_1$,
$K_2$ and $K_3=i[K_1, K_2]/2$ satisfy an $SL(2,R)$ algebra.
This was also shown to be
the case for the nonabelian models in section 6 for
appropriate definitions of $K_1$ and
$K_2$. Although this was not true for the most
natural generalization of $K_1$ and
$K_2$ even
these operators were  shown to
satisfy effectively an $SL(2,R)$ algebra, \ie they
satisfy an $SL(2,R)$ algebra on the states $|\phi\hb_l$.
Consequently the factor $e^{[Q, \psi]}$ may be
viewed as a group transformation belonging to a
one-dimensional subgroup of $SL(2, R)$.
When $\al$ and $\beta$ have the same signs it belongs to a compact subgroup
while
opposite signs of $\al$ and $\beta$ makes it belong to a noncompact one. These
two
possibilities are quite different. In fact, there is a strong argument against
the
first possibility. One may notice that the connection between the
representation
\r{705} and the gauge fixing in the conventional BFV theory requires us
in fact to
identify our $[Q, \psi]$ with $tH$ where $H$ is a Hamiltonian operator
given by $[Q,
\psi']$ \cite{Path}. The proper identification of $\psi$ and $\psi'$ is
therefore $\psi=t\psi'$. The replacement of $\al$, $\beta$ by $t\al$,
$t\beta$ in \r{706} leads then to \r{7061} with \be
&&\al'=\al\,\frac{\tan{(|t|\sqrt{\al\beta})}}{\sqrt{\al\beta}}\sign\, t,
\;\;\;\beta'=\beta\,\frac{\tan{(|t|\sqrt{\al\beta})}}{\sqrt{\al\beta}}\sign\, t
\e{7081} for $\al\beta>0$ and \be
&&\al'=\al\,\frac{\tanh{(|t|\sqrt{-\al\beta})}}{\sqrt{-\al\beta}}\sign\, t,
\;\;\;\beta'=
\beta\,\frac{\tanh{(|t|\sqrt{-\al\beta})}}{\sqrt{-\al\beta}}\sign \,t
\e{7082} for $\al\beta<0$.
Thus, if $\al$ and $\beta$ have the same sign, $|s\hb_1$ and $|s\hb_2$ in
\r{7061}
will be badly defined for infinitely many instants, $t=n\pi/(2\sqrt{\al\beta})$
where
$n$ is an integer, while opposite signs of $\al$ and $\beta$ makes $t=0$ the
only
badly defined instant. Remarkably enough the coBRST charge \r{7041}  belongs
to the
latter category and is therefore a good gauge fixing fermion even in this more
restricted sense. In fact, our analysis indicates that  any
noncompact gauge choice ($\al\beta<0$) may be represented by a coBRST charge.

{}From our results  in sections 5 and 6
it is also  possible to make use of  a more general $^*Q$ than
\r{7041} in \r{707}. In the general case
the form \r{7042} should be relevant as a gauge fixing fermion provided  the
gauge
fixing variables $\chi^a$ and $\Lambda^a$ are abelian. This
 form of $\psi$ we have in \eg QED and Yang-Mills. One may
notice that essentially only abelian gauge fixing has been used in the
literature so
far. (This is \eg required in the proof of gauge invariance given in
\cite{Fad}.) It
would  certainly be interesting to understand what possible role the
nonlinear terms
in the coBRST charge $^*Q$ for nonabelian gauge fixing can possibly play when
$^*Q$ is viewed as a gauge fixing fermion. Anyway, apart from this question
mark, our
results   suggest that the coBRST charge $^*Q$ is always a good gauge fixing
fermion
and  a candidate for a natural choice of $\psi$.

We end with a comment on the difference between coBRST and antiBRST. These two
concepts are often confused in the literature. Like the coBRST charge ($^*Q$)
also
the antiBRST charge ($\bar{Q}$) has ghost number minus one and is nilpotent.
However,
contrary to the coBRST charge the antiBRST charge anticommutes with the BRST
charge
and is a symmetry of the model. For the simple abelian model  in section 2 the
antiBRST charge is given by \cite{CFO,SH2}
\be\bar{Q}=p_a\baca^a-\pet_a\pi^a=[Q, \pet_a\baca^a]
\e{708}
In this case the coBRST charge \r{7041} may be expressed in terms of
$\bar{Q}$ as
follows \be
^*Q=i[\bar{Q}, S]
\e{709}
where
\be
S\equiv
\frac{\la}{2}\left(rx_ax^a+\frac{1}{r}v_a v^a\right)
\e{710}
The form \r{709} of $^*Q$ is also the form of a gauge fixing fermion in
antiBRST
invariant theories (cf. \cite{SH2}).

There are also other charges in the literature which have ghost number
minus one and
which are nilpotent. The proposal of coBRST in \cite{Solv} for the abelian
model was
$Q'=\baca_a v^a+\pet_ax^a$ which differs from \r{705} (it yields zero on
$|\phi\hb_l$). In \cite{JLG,BY,HH}  a $Q'$ is defined by exchanging all ghosts
with
their conjugate momenta which implies $Q'=p_a\pet^a+\baca_a\pi^a$ for the
abelian
model. (This $Q'$ was called antiBRST in \cite{BY} and coBRST in \cite{HH}.)

\setcounter{section}{1}
\setcounter{equation}{0}
\renewcommand{\thesection}{\Alph{section}}
\newpage
\noindent
{\Large{\bf{Appendix A}}}
 \vspace{5mm}\\
{\bf Proof of eq.\r{316}}\\ \\

Making use of formulas (B.20)-(B.30) in appendix B of ref.\cite{Quaade} we
find the
following equalities
\be
&&e^{\al K_1+\beta K_2}=e^{\ga K_2}e^{\del K_1}e^{\ga K_2}
=e^{\ga' K_1}e^{\del' K_2}e^{\ga' K_1}
\e{a11}
where the parameters $\ga,\del,\ga'$ and $\del'$ are given by
\be
&&\ga=\frac{\sqrt{\al\beta}}{\al}\tan\halv{\sqrt{\al\beta}},\;\;\;
\del=\frac{\al}{\sqrt{\al\beta}}\,\sin{\sqrt{\al\beta}}\nn\\
&&\ga'=\frac{\sqrt{\al\beta}}{\beta}\tan\halv{\sqrt{\al\beta}},\;\;\;
\del'=\frac{\beta}{\sqrt{\al\beta}}\,\sin{\sqrt{\al\beta}}\nn\\
&&\cos{\sqrt{\al\beta}}=1-\del\ga=1-\del'\ga'
\e{a12}
for $\al\beta>0$. For $\al\beta<0$ we have the same relations with the
replacement
\r{37}.

By means of \r{a11} and
\be
&&K_2|\phi\hb_1=K_1  |\phi\hb_2=K_3    |\phi\hb_1=K_3|\phi\hb_2=0
\e{a13}
 we get therefore
\be
&&|s\hb_1=e^{\al K_1+\beta K_2}|\phi\hb_1=e^{\ga K_2}e^{\del
K_1}|\phi\hb_1=
\sum_{n,m=0}^\infty\frac{\ga^n\del^m}{n!m!}K_2^nK_1^m|\phi\hb_1,\nn\\
&&|s\hb_2=e^{\al K_1+\beta K_2}|\phi\hb_2=e^{\ga'
K_1}e^{\del'
K_2}|\phi\hb_2=
\sum_{n,m=0}^\infty\frac{\ga^{'n}\del^{'m}}{n!m!}K_1^nK_2^m|\phi\hb_2
\e{a14}
It is easily seen from the algebra \r{314} that
\be
&&[K_1, K_2^n]=-2inK_2^{n-1}K_3+n(n-1)K_2^{n-1},\nn\\
&&[K_2, K_1^n]=2inK_1^{n-1}K_3+n(n-1)K_1^{n-1}
\e{a15}
Hence, due to \r{a13} we have
\be
K_2K_1^n|\phi\hb_1=n(n-1)K_1^{n-1}|\phi\hb_1,\;\;\;
K_1K_2^n|\phi\hb_2=n(n-1)K_2^{n-1}|\phi\hb_2
\e{a16}
which implies
\be
&&K_2^mK_1^n|\phi\hb_1=
n(n-1)^2(n-2)^2\cdots(n-m+1)^2(n-m)K_1^{n-m}|\phi\hb_1,\nn\\
&&K_1^mK_2^n|\phi\hb_2=n(n-1)^2(n-2)^2\cdots(n-m+1)^2(n-m)K_2^{n-m}|\phi\hb_2
\e{a17}
When this is inserted into \r{a14} we find therefore
\be
&&|s\hb_1=\exp{\left(\frac{\del}{1-\del\ga}
K_1\right)}|\phi\hb_1,\;\;\;|s\hb_2=\exp{\left(\frac{\del'}{1-\del'\ga'}
K_2\right)}|\phi\hb_2
\e{a18}
where
\be
\frac{\del}{1-\del\ga}=\frac{\al}{\sqrt{\al\beta}}\,\tan{\sqrt{\al\beta}},
\;\;\;\frac{\del'}{1-\del'\ga'}=
\frac{\beta}{\sqrt{\al\beta}}\,\tan{\sqrt{\al\beta}}
\e{a19}

\setcounter{section}{2}
\setcounter{equation}{0}
\newpage
\noindent
{\Large{\bf{Appendix B}}}
 \vspace{5mm}\\
{\bf Some unitary symmetries.}\\ \\
Let us introduce the following unitary operators for the abelian model
introduced in
section~2
 \be
&&U_1(\ga)\equiv\exp{\{i\ga\halv(p_ax^a+x^ap_a)\}},\;\;\;
U_2(\ga)\equiv\exp{\{i\ga\halv(\pi_av^a+v^a\pi_a)\}}\nn\\
&&U_3(\ga)\equiv\exp{\{\ga\halv(\ca^a\pet_a-\pet_a\ca^a)\}},\;\;\;
U_4(\ga)\equiv\exp{\{\ga\halv(\baca_a\bapet^a-\bapet^a\baca_a)\}}
\e{a1}
where $\ga$ is a real constant. These operators act as scaling operators on the
original variables. The nontrivial transformations are
 \be
&&U_1(\ga)x^aU_1\dagg(\ga)=e^\ga x^a,\;\;\;U_1(\ga)p_aU_1\dagg(\ga)=
e^{-\ga} p_a\nn\\
&&U_2(\ga)v^aU_2\dagg(\ga)=e^\ga v^a,\;\;\;U_2(\ga)\pi_aU_2\dagg(\ga)=e^{-\ga}
\pi_a\nn\\
&&U_3(\ga)\ca^aU_3\dagg(\ga)=e^\ga \ca^a,\;\;\;U_3(\ga)\pet_aU_3\dagg(\ga)=
e^{-\ga}
\pet_a\nn\\ &&U_4(\ga)\baca_aU_4\dagg(\ga)=e^\ga
\baca_a,\;\;\;U_4(\ga)\bapet^aU_4\dagg(\ga)=e^{-\ga} \bapet^a
 \e{a2}
Consider then the representation
\be
&&|s\hb=e^{[Q, \psi]}|\phi\hb
\e{a3}
where $|\phi\hb$ satisfies the properties of $|\phi\hb_1$ or $|\phi\hb_2$ in
\r{26}.
Under the unitary transformations \r{a1} they satisfy ($a=1,2,\ldots,n$)
\be
&&U_1(\ga)|\phi\hb_1=e^{-\halv
n\ga}|\phi\hb_1,\;\;\;U_2(\ga)|\phi\hb_1=e^{\halv
n\ga}|\phi\hb_1\nn\\
&&U_3(\ga)|\phi\hb_1=e^{\halv
n\ga}|\phi\hb_1,\;\;\;U_4(\ga)|\phi\hb_1=e^{\halv
n\ga}|\phi\hb_1\nn\\
&&U_1(\ga)|\phi\hb_2=e^{\halv
n\ga}|\phi\hb_2,\;\;\;U_2(\ga)|\phi\hb_2=e^{-\halv
n\ga}|\phi\hb_2\nn\\
&&U_3(\ga)|\phi\hb_2=e^{-\halv
n\ga}|\phi\hb_2,\;\;\;U_4(\ga)|\phi\hb_2=e^{-\halv
n\ga}|\phi\hb_2
 \e{a4}
which is another sign of the fact  that $|\phi\hb$ is {\em not} an inner
product
state. We notice now that the following combinations of the unitary operators
\r{a1} leave the BRST charge \r{21} invariant:
\be
&&U_1(\ga)U_3(\ga),\;\;\;U_2(\ga)U_4\dagg(\ga),\;\;\;
U_1(\ga)U_2(\ga)U_3(\ga)U_4\dagg(\ga)\nn\\
&&U_1(\ga)U_2\dagg(\ga)U_3(\ga)U_4(\ga)
\e{a5}
and the following combinations scale $Q$:
\be
&&U_1(\ga)U_2(\ga),\;\;\;U_3\dagg(\ga)U_4(\ga),\;\;\;U_1(\ga)U_4(\ga),
\;\;\;U_2(\ga)U_3\dagg(\ga)\nn\\
&&U_1(\ga)U_2(\ga)U_3\dagg(\ga)U_4(\ga)
\e{a6}
(The first four combinations yield $Q\,\ra\,e^{-\ga}Q$ and the last
$Q\,\ra\,e^{-2\ga}Q$.)  All  combinations in \r{a5} and \r{a6} yield unity on
$|\phi\hb$ for any of the two sets of conditions in \r{26}.

Let now $U$ be any of the combinations in \r{a5} and \r{a6}. We find then for
the
representation \r{a3} with $\psi=\al\pet_av^a+\beta\baca_a x^a$:
 \be
&&U|s\hb=e^{[Q,\psi']}|\phi\hb
\e{a7}
where
\be
&&\psi'=\psi \mbox{ \ for \ }
U=U_1(\ga)U_2(\ga),\;U_3(\ga)U_4\dagg(\ga)\nn\\
&&\psi'=\al e^{-\ga}\pet_av^a+\beta e^{\ga}\baca_a x^a \mbox{ \ for \ }
U=U_1(\ga)U_4(\ga),\;U_2\dagg(\ga) U_3(\ga)
\e{a8}
Notice that the generator of  $U_3(\ga)U_4\dagg(\ga)$ ($U_1(\ga)U_2(\ga)$) is
the
ghost number operator when we have fermionic (bosonic) ghosts.  $|s\hb$
is  invariant under these combinations since
$|s\hb$ has ghost number zero.

For the nonabelian models of section 6 we still have that
$U_2(\ga)U_4\dagg(\ga)$
leaves the BRST charge \r{601} invariant, and that $U_3\dagg(\ga)U_4(\ga)$ and
$U_2(\ga)U_3\dagg(\ga)$ scale it ($Q\,\ra\,e^{-\ga}Q$). Thus,  \r{a7}
is still valid with $\psi'=\psi$ for $U_3(\ga)U_4\dagg(\ga)$ and
$\psi'=\al e^{-\ga}\pet_av^a+\beta e^{\ga}\baca_a x^a$ for
$U_2\dagg(\ga) U_3(\ga)$.
Notice that the ghost number operator is also here
the generator of $U_3(\ga)U_4\dagg(\ga)$.\\
\vspace{1cm}\\

\setcounter{section}{3}
\setcounter{equation}{0}
\noindent
{\Large{\bf{Appendix C}}}
 \vspace{5mm}\\
{\bf Some properties used in section 6.}\\ \\
In section 6 the gauge fixing variables $x^a$ satisfy
\be
&&[x^a, x^b]=0,\;\;\;[x^a, \theta_b]=iM^a_{\;\;b}
\e{c1}
where $M^a_{\;\;b}$ is an invertible matrix operator. If $x^a$ are chosen to be
canonical group coordinates then $M^a_{\;\;b}$  depends only on $x^a$ and
satisfies
the equations \be
&&(\partial_d
M^c_{\;\;a})M^d_{\;\;b}-(\partial_d
M^c_{\;\;b})M^d_{\;\;a}=U_{ab}^{\;\;\;d}M^c_{\;\;d}  \e{c2}
where the derivatives are with respect to $x^a$. These are the equations
for the vielbeins of the group. The solution may be obtained as a power
series in $x^a$. To the first
orders we have
\be
&M^a_b = &{\delta}^a_b + \frac{1}{2} U_{bc}^{\;\;a} x^c +
\frac{1}{12} U_{be_1}^{\;\;d} U_{de_2}^{\;\;a} x^{e_1} x^{e_2}
- \nn\\&&-\frac{1}{720} U_{be_1}^{\;\;d_1} U_{d_1 e_2}^{\;\;\;d_2}
U_{d_2 e_3}^{\;\;\;d_3} U_{d_3 e_4}^{\;\;a}
x^{e_1} x^{e_2} x^{e_3} x^{e_4} + O(x^6)
\e{c3}
The inverse of $M$ is of the particularly simple form:
\be
&(M^{-1})^a_{\;\;b} =& {\delta}^a_b
- \frac{1}{2!} U_{bc}^{\;\;a} x^c +
\frac{1}{3!} U_{be_1}^{\;\;d} U_{de_2}^{\;\;a} x^{e_1} x^{e_2}
- \nn\\&&-\frac{1}{4!} U_{be_1}^{\;\;\;d_1} U_{d_1 e_2}^{\;\;\;d_2}
U_{d_2 e_3}^{\;\;\;a} x^{e_1} x^{e_2} x^{e_3}+\nn\\
&&+ \frac{1}{5!} U_{be_1}^{\;\;\;d_1} U_{d_1 e_2}^{\;\;\;d_2} U_{d_2
e_3}^{\;\;\;d_3}
U_{d_3 e_4}^{\;\;\;a}
x^{e_1} x^{e_2} x^{e_3} x^{e_4} + O(x^5)
\e{c31}
In terms of this $M^a_{\;\;b}$ the hermitian gauge generators $\theta_a$ may
be
represented as  \be
&&\theta_a=\halv\left(p_bM^b_{\;\;a} + M^b_{\;\;a}p_b  \right)
\e{c4}
where $p_a$ are hermitian conjugate momenta to $x^a$.
{}From \r{c1} and \r{c4}  we have
\be
&&p_a=(M^{-1})^b_{\;\;a}\theta_b+i\halv
(M^{-1})^b_{\;\;a}\partial_cM^c_{\;\;b},\;\;\;[x^a,
p_b]=i\del^a_b \e{c5}
Notice that \r{c2} implies $[\theta_a, \theta_b]=iU_{ab}^{\;\;\;c}\theta_c$.
Another
property of $M^a_{\;\;b}$ is
\be
&&M^a_{\;\;b}x^b=(M^{-1})^a_{\;\;b}x^b=x^a
\e{c6}\\ \\
{\bf Proof of \r{614}:}\\
Using the form \r{610} of $K_1$ and $K_2$ as well as the definition \r{611}
of $K_3$ we
have \be
&&[K_2, K_3]
=-i\pi_a M^a_{\;\;b} x^b+\baca_aM^a_{\;\;b}M^b_{\;\;c}\ca^c+i\baca_a\ca^b
x^c[\theta_b, M^a_{\;\;c}]
\e{c7}
where we have made use of the general Jacobi identities \r{605}.
Now from \r{c6} and
\be
&&M^a_{\;\;b}=i[\theta_b, M^a_{\;\;c}]x^c+M^a_{\;\;c}M^c_{\;\;b}
\e{c8}
which follows from \r{c6} (use $[\theta_b, M^a_{\;\;c}x^c-x^a]=0$) we
find
\be
&&[K_2, K_3]=iK_2
\e{c9}

Similarly we find straight-forwardly
\be
&&[K_1, K_3]=i(\theta_a+\theta_a^{gh})\left(M^a_{\;\;b}+\halv
U_{cb}^{\;\;\;a}x^c\right)v^b -\pet_a\left(M^a_{\;\;b}+\halv
U_{cb}^{\;\;\;a}x^c\right)v^b+\nn\\ &&+\halv\pi_a[\theta_d, M^a_{\;\;b}]v^bv^d
+\pet_a\ca^b v^c\left(\halv U_{cd}^{\;\;\;a}M^d_{\;\;b}-i[\theta_b,
M^a_{\;\;c}]\right)+\nn\\&& +i\halv\baca_a\bapet^b v^c\left([\theta_b,
M^a_{\;\;c}]+[\theta_c, M^a_{\;\;b}]\right) -
\halv\baca_a\ca^bv^cv^d[\theta_b,[\theta_d,
M^a_{\;\;c}]] \e{c10}
where we have made use of the general Jacobi identities \r{612} and
\be
&&U_{db}^{\;\;\;e}[\theta_e, M^a_{\;\;c}]=
i[\theta_b, [\theta_d,  M^a_{\;\;c} ]]-
i[\theta_d, [\theta_b,  M^a_{\;\;c} ]]
\e{c11}
We notice now that if $x^a$ are chosen as canonical coordinates on the group
manifold
then we have
\be
&&\left(M^a_{\;\;b}+\halv
U_{cb}^{\;\;\;a}x^c\right)|\phi\hb_1=\del^a_b|\phi\hb_1,\;\;\; \left([\theta_b,
M^a_{\;\;b}]+[\theta_c, M^a_{\;\;b}]\right)|\phi\hb_1=0 \e{c12}
and
\be
&&[K_1, K_3]|\phi\hb_1=-iK_1|\phi\hb_1
\e{c13}
This proves eq.\r{614}$\bullet$

Eq.\r{c9} is easily seen to imply
\be
&&[K_1, K_2^n]=-2inK_2^{n-1}K_3+n(n-1)K_2^{n-1}
\e{c14}
Hence,  we have
\be
&&K_1K_2^n|\phi\hb_2=n(n-1)K_2^{n-1}|\phi\hb_2
\e{c15}
We also expect
\be
&&K_2K_1^n|\phi\hb_1=n(n-1)K_1^{n-1}|\phi\hb_1
\e{c16}
This requires apart from \r{c13} also
\be
&&[K_1,[K_1,K_3]]|\phi\hb_1=[K_1,[K_1,[K_1,K_3]]]|\phi\hb_1=\ldots=0
\e{c17}
This we have checked  to lowest order and the structure of \r{c10} seems to
make
it
true for any order. However, we have no rigorous proof. Anyway we feel  rather
confident that \r{c16} is valid for any $n$.  Eqs. \r{c15} and \r{c16} imply
\r{a17} and the formulas \r{a18} in appendix A.  (Notice that the formulas
\r{a11} were
not necessary for the derivation of \r{a18}. They only provided for a
convenient way
to obtain the nice expressions of the coefficients in the expansions \r{a14}.)

\end{document}